\documentclass[11pt,a4paper]{article}
\usepackage{epsfig}
\usepackage{a4wide}
\usepackage{latexsym}
\usepackage{graphicx}
\usepackage{amsmath}
\usepackage{slashed}
\usepackage{amsfonts}
\usepackage{amssymb}
\usepackage{mathrsfs}
\usepackage{appendix}
\usepackage[usenames,dvips]{color}
\usepackage[colorlinks=true,urlcolor=PineGreen,anchorcolor=PineGreen,citecolor=PineGreen,filecolor=PineGreen,linkcolor=PineGreen,menucolor=PineGreen,pagecolor=PineGreen,linktocpage=true,pdfproducer=medialab]{hyperref}
\usepackage{layout}

\allowdisplaybreaks[4]
\numberwithin{equation}{section}

\begin{document}

\begin{titlepage}
\begin{flushright}

\end{flushright}

\begin{center}
{\Large {\textbf{{Special Galileon at one loop}}}}\\[1.5 cm]%
\textbf{Filip P\v{r}eu\v{c}il and Ji\v{r}\'{\i} Novotn\'y}\footnote{
for emails use: \textit{surname\/} at ipnp.troja.mff.cuni.cz},  \\[1 cm]
\textit{Institute of Particle and Nuclear Physics, Faculty of
Mathematics
and Physics,}\\[0pt]
\textit{Charles University, V Hole\v{s}ovi\v{c}k\'ach 2, CZ-180 00
Prague 8,
Czech Republic} \\[0.5 cm]
\end{center}

\begin{abstract}
We present a complete one-loop renormalization of the Special Galileon $S-$matrix. Especially we give a complete list of the higher derivative operators which are necessary for one-loop on-shell renormalization and  prove the invariance of the one-loop on-shell effective action with respect to the Special Galileon symmetry. This enables us to enlarge the validity of the enhanced $O(p^3)$ soft behavior of the scattering amplitudes to the one-loop level.  As an illustration we discuss explicitly the four-point and five-point one-loop scattering amplitudes and comment on some conjectures appearing in the existing literature. 

\end{abstract}

\end{titlepage}

\tableofcontents

\bigskip \line(1,0) {100}

\setcounter{footnote}{0} 

\section{Introduction: Why is the Special Galileon special?}

The Galileons are remarkable derivatively coupled scalar effective field
theories with plethora of very interesting properties both at the classical
and the quantum level. In the literature, the (cubic) Galileon emerged at
first time as the only interacting (zero helicity) mode in the decoupling
limit of the Dvali-Gababadze-Poratti modified gravity model \cite%
{Dvali:2000hr,Deffayet:2001pu} and almost at the same time also in the
similar limit of the massive gravity model \cite{deRham:2010kj}. Soon it has
been recognized \cite{Nicolis:2008in} that the Galileon in its generalized
form might be a promising local modification of the General relativity at
large scales with several appealing properties. In particular, regardless of
the presence of highly nonlinear higher derivative couplings in the
Lagrangian, the Galileon obeys the second order equation of motion which
ensures the absence of the Ostrogradsky ghosts. It has been also shown, that
near the massive sources the Galileon field is suppressed by the Vainshtein
screening mechanism \cite{Vainshtein:1972sx} and that the basic Lagrangian
is stable with respect to the quantum corrections \cite%
{Luty:2003vm,Hinterbichler:2010xn,deRham:2012ew}. The seminal paper \cite%
{Nicolis:2008in} initialized a boom of increasing interest in the Galileon
theories concerning their formal properties , the possible generalizations,
e.g. \cite{Deffayet:2009mn,Deffayet:2010zh} as well as the cosmological
applications. For a pedagogical reviews and for more comprehensive list
existing literature see e.g. \cite{Curtright:2012gx, Khoury:2013tda}

In the general case, the basic flat space Galileon Lagrangian collects the
most general terms built of the single scalar field and its derivatives (up
to and including the second order ones, with $n$ fields and $2n-2$
derivatives). These terms can be uniquely determined by the requirement of
yielding the second order equations of motion and by invariance with respect
to the polynomial shift symmetry of the first order,%
\begin{equation}
\delta \phi =a+b\cdot x,  \label{general Galileon symmetry}
\end{equation}%
where $a$ and $b_{\mu }$ are real parameters. The latter property has an
important consequence at the quantum level, namely the tree-level on-shell
scattering amplitudes poses an enhanced soft behavior \cite%
{Cheung:2014dqa,Cheung:2015ota,Cheung:2016drk}. This means that the
scattering amplitudes vanish as the second power of momentum when one of the
external particles becomes soft, which is higher than one would naively
expect from the simple counting of the derivatives in the Lagrangian. The
general basic Galileon Lagrangian in $D$ dimensions can be written in the
form%
\begin{equation}
\mathcal{L}_{b}=\sum_{n=0}^{D}d_{n+1}\phi \varepsilon ^{\mu _{1}\ldots \mu
_{D}}\varepsilon ^{\nu _{1}\ldots \nu _{D}}\prod_{i=1}^{n}\partial _{\mu
_{i}}\partial _{\nu _{i}}\phi \prod_{j=n+1}^{D}\eta _{\mu _{j}\nu _{j}}
\label{lagrangian}
\end{equation}%
where $d_{n}$ are free real $n-$point couplings. For the theory to be well
defined on the quantum level we demand $d_{1}=0$ (no tadpoles) and $%
d_{2}=\left( -1\right) ^{D}/2\left( D-1\right) !$ (canonical normalization
of the kinetic term). The above mentioned enhanced soft behavior enables the
full on-shell reconstructibility of the tree-level scattering amplitudes
once the basic set of the seed amplitudes from the four-point up to the $D+1$
point one is known \cite{Cheung:2015ota}. On the other hand, due to the rich
set of dualities of the Galileon Lagrangian \cite{deRham:2013hsa,
deRham:2014lqa,Creminelli:2014zxa,Kampf:2014rka}, there is a many-to-one
correspondence between the constants $d_{n}$ and the on-shell physics
represented by the scattering amplitudes.

Particular choice of the couplings $d_{n}$ can further increase the symmetry
of the Lagrangian (\ref{lagrangian}). Namely, we mean the choices 
\begin{equation}
d_{2n}^{\left( +\right) }=\frac{\left( -1\right) ^{n}}{2n}\left( 
\begin{array}{c}
D \\ 
2n-1%
\end{array}%
\right) \frac{\cos \beta }{\alpha ^{2(n-1)}},~~~~~d_{2n+1}^{\left( +\right)
}=\frac{\left( -1\right) ^{n}}{2n+1}\left( 
\begin{array}{c}
D \\ 
2n%
\end{array}%
\right) \frac{\sin \beta }{\alpha ^{2n-1}},  \label{sGal1}
\end{equation}%
or%
\begin{equation}
d_{2n}^{\left( -\right) }=\frac{1}{2n}\left( 
\begin{array}{c}
D \\ 
2n-1%
\end{array}%
\right) \frac{\cosh \beta }{\alpha ^{2(n-1)}},~~~~~d_{2n}^{\left( -\right) }=%
\frac{1}{2n+1}\left( 
\begin{array}{c}
D \\ 
2n%
\end{array}%
\right) \frac{\sinh \beta }{\alpha ^{2n-1}}.  \label{sGal2}
\end{equation}%
Here $\alpha $ has dimension $\left[ mass\right] ^{(D+2)/2}$ and $\alpha
^{2/(D+2)}$ is therefore the scale which controls the size of nonlinearities
in the Lagrangian (\ref{lagrangian}) while $\beta $ is dimensionless
parameter. The prescription (\ref{sGal1}) and (\ref{sGal2}) yield two
branches of two parametric families of Lagrangians $\mathcal{L}^{\left( \pm
\right) }\left( \alpha ,\beta \right) $ which are invariant with respect to
the generalized polynomial shift symmetry \cite{Hinterbichler:2015pqa,
Novotny:2016jkh}%
\begin{equation}
\delta \phi =\frac{\theta }{2}H^{\mu \nu }\left( \alpha ^{2}x_{\mu }x_{\nu
}\pm \partial _{\mu }\phi \partial _{\nu }\phi \right)
\label{sGal_transform}
\end{equation}%
where $H^{\mu \nu }$ is arbitrary fixed traceless\footnote{%
For $H_{\mu }^{\mu }\neq 0$ the transformation (\ref{sGal_transform}) is a
duality of the family, which transforms Lagrangian (\ref{lagrangian}) and (%
\ref{sGal2}) with parametres $\left( \alpha ,\beta \right) $ into the
Lagrangian with parametres $\left( \alpha ,\beta \mp \theta \alpha H_{\mu
}^{\mu }\right) $.} symmetric tensor and $\theta $ is an infinitesimal
parameter. On the quantum level this symmetry is responsible for further
enhancement of the soft behavior. In the above theories with $\beta =0$ the
tree-level amplitudes vanish as the third power of momentum in the single
particle soft limit. The latter theories (\ref{sGal1}) and (\ref{sGal2})
with $\beta =0$ are known as Special Galileons \cite%
{Cheung:2014dqa,Cachazo:2014xea}.

The Special Galileons are special in several aspects. First, their soft
behavior is in a sense extremal, since as it was shown in \cite%
{Cheung:2016drk}, the $O\left( p^{3}\right) $ single soft limit is the
highest possible within the single scalar effective theories with nontrivial
power counting. To be more precise, let us assume a derivatively coupled
single scalar effective field theory and for each elementary vertex $V$ in
the Lagrangian let us determine the ratio 
\begin{equation*}
\rho _{V}=\frac{D_{V}-2}{E_{V}-2},
\end{equation*}%
where $D_{V}$ is a number of derivatives and $E_{V}$ is a number of external
legs of the vertex $V$. The soft behavior of such a theory can be
characterized by the soft exponent $\sigma $ which corresponds to the $%
O\left( p^{\sigma }\right) $ behavior of the amplitudes in the single
particle soft limit $p\rightarrow 0$. Then for the general Galileon we get $%
\rho _{V}=2$ for each admissible vertex and the soft exponent is $\sigma =2$%
. Provided we assume only theories with fixed $\rho _{V}=\rho \geq 1$, i.e.
only with vertices with $2+\rho \left( E_{V}-2\right) $ derivatives and with
the soft limit characterized by soft exponent $\sigma $, then the theories
with nontrivial enhanced soft behavior are those for which $\rho \leq \sigma 
$ since in the opposite case there is enough derivatives per field in each
vertex to ensure the $O\left( p^{\sigma }\right) $ soft limit automatically.
However, as shown in \cite{Cheung:2016drk}, not all the nontrivial pairs $%
\left( \rho ,\sigma \right) $ are admissible, namely they have to satisfy
the bounds $\rho \geq \sigma -1$ and $\sigma \leq 3$. Note that the Special
Galileons saturate both this bounds sitting in the very corner of the
admissible $\left( \rho ,\sigma \right) $ region. This means that the
Special Galileon together with the Dirac-Born-Infeld theory and the
Non-linear sigma model belongs to the set of the exceptional scalar
effective field theories for which there exist the Cachazo-He-Yuan
representation \cite{Cachazo:2014xea}.

The second special property of the Special Galileon is that it is (contrary
to the general Galileon theories) potentially well defined as an
perturbative effective quantum field theory - at least formally when we
assume that the regularization and renormalization preserves all the
symmetries. The full Lagrangian is then 
\begin{equation}
\mathcal{L}=\mathcal{L}_{b}+\mathcal{L}_{CT}
\end{equation}%
where $\mathcal{L}_{CT}$ corresponds to the higher derivative counterterms
needed for perturbative renormalization. Here we tacitly assume that $%
\mathcal{L}_{CT}$ shares the symmetries of the basic Lagrangian $\mathcal{L}%
_{b}$. As was shown in \cite{Kampf:2014rka}, it is possible to organize the
perturbative expansion according to the hierarchy of the vertices and
according to the hierarchy of the corresponding Feynman graphs. This
hierarchy is based on the assignment of an index $i_{V}$ to each vertex $V$
(including those coming from $\mathcal{L}_{CT}$), where%
\begin{equation}
i_{V}=D_{V}-\left( 2E_{V}-2\right) ,  \label{index of the vertex}
\end{equation}%
Here $D_{V}$ and $E_{V}$ have the same meaning as above. Note that for the
basic Lagrangian (\ref{lagrangian}) we get $i_{V}=0$ and therefore this
index measures the abundance of the number of derivatives relatively to the
basic Lagrangian $\mathcal{L}_{b}$. The index of the counterterm vertices
which are necessary in order to renormalize the one particle irreducible
graph $\Gamma $ is then given (for derivation of theis formula see \cite%
{Kampf:2014rka})%
\begin{equation}
i_{CT}^{\Gamma }=(D+2)L_{\Gamma }+\sum\limits_{V\in \Gamma }i_{V}
\label{index of graph}
\end{equation}%
where $L_{\Gamma }$ is number of loops and $i_{V}$ are the indices of the
vertices of the graph $\Gamma $. The indices of the counterterms are thus
related to the loop expansion. Therefore, in principle, the perturbative
renormalization in the effective theory sense would be possible provided
there is only finite number of vertices with fixed $i_{CT}$. Then at each
level of the hierarchy, given by the contributions of all the Feynman graphs 
$\Gamma $ with $i_{CT}^{\Gamma }$ fixed, there would be only finite number
of unknown parameters.

In the case of the general Galileon, the only other constraint for the
counterterms is the invariance with respect to the symmetry transformation (%
\ref{general Galileon symmetry}). Then, however, there is an infinite number
of possible uncorrelated vertices with given fixed $i_{CT}$ and therefore
the formal effective theory looses its predictivity at each hierarchy level
(see \cite{Kampf:2014rka} for more detailed discussion).

On the other hand, the generalized polynomial shift symmetry (\ref%
{sGal_transform}) of the Special Galileon is strong enough to restrict the
form of the counterterm Lagrangian with $i_{CT}$ fixed in such a way, that
at each hierarchy level there is only finite number of unknown parameters in
the $S-$matrix. The situation here is somewhat similar to the case of chiral
perturbation theory \cite{Gasser:1983yg, Gasser:1984gg} where the
nonlinearly realized chiral symmetry allows only for finite number of low
energy couplings at each order in the derivative expansion. This of course
does not mean that there is only finite number of vertices at each hierarchy
level. Rather the infinite number of possible vertices have correlated
couplings and as a result they are combined into a finite number of
operators invariant with respect to (\ref{sGal_transform}).

The third distinguished feature of the Special Galileon is the beautiful
geometry behind the polynomial shift symmetry (\ref{sGal_transform}). The
Special Galileon field can be understood as a scalar degree of freedom which
describes fluctuations of a $D-$dimensional brane in a $2D-$dimensional
pseudo-Riemanian target space $%
\mathbb{R}
^{2,2D-2}$ treated as a K\={a}hler manifold\footnote{%
This is the case of the \textquotedblleft plus\textquotedblright\ branch of
the Special Galileon. For the \textquotedblleft minus\textquotedblright\
branch an appropriate analytic continuation of the parameter $\alpha $ is
necessary. The target space is then $%
\mathbb{R}
^{D,D}$ which has no compatible complex structure. See for \cite%
{Novotny:2016jkh} more details.}. The hidden Special Galileon symmetry (\ref%
{sGal_transform}) corresponds then to the nonlinearly realized subgroup of
the symmetry group $%
\mathbb{C}
^{D}\rtimes U\left( 1,D-1\right) $ of the target space. This interpretation
of the Special Galileon allows for simple construction and classification of
the counterterm Lagrangian $\mathcal{L}_{CT}$ (see \cite{Novotny:2016jkh}
for more details)\footnote{%
Alternative way of the classification of the higher derivative Special
Galileon Lagrangians was developed in \cite%
{Garcia-Saenz:2019yok,Carrillo-Gonzalez:2019aao} using the coset
construction.}. The basic building blocks for higher derivative Lagrangians
are then the effective metric (the $\pm $ in the following formulas
correspond to the two choices of the transformation prescription (\ref%
{sGal_transform})) 
\begin{equation}
g_{\mu \nu }=\eta _{\mu \nu }\pm \frac{1}{\alpha ^{2}}\partial _{\mu
}\partial \phi \cdot \partial \partial _{\nu }\phi ,  \label{aGal metric}
\end{equation}%
the extrinsic curvature tensor%
\begin{equation}
\mathcal{K}_{\alpha \mu \nu }=-\frac{1}{\alpha }\partial _{\alpha }\partial
_{\mu }\partial _{\nu }\phi  \label{extrinsic curvature}
\end{equation}%
the Christoffell symbol 
\begin{equation}
\Gamma _{\rho \sigma \mu }=\pm \frac{1}{\alpha ^{2}}\partial _{\mu }\partial
_{\sigma }\partial \phi \cdot \partial \partial _{\rho }\phi ,
\label{christoffell symbol}
\end{equation}%
the corresponding covariant derivative $D_{\mu }$ and the covariant
Levi-Civita tensor%
\begin{equation}
E^{\mu _{1}\ldots \mu _{D}}=\frac{1}{\sqrt{g}}\varepsilon ^{\mu _{1}\ldots
\mu _{D}}  \label{levi-civita}
\end{equation}%
where $g$ is the absolute value of the determinant of the effective metric $%
g_{\mu \nu }$. Any diffeomorphism invariant built from $g_{\mu \nu }$ and
its inverse $g^{\mu \nu }$, ${\mathcal{K}}_{\alpha \mu \nu }$, and their
covariant derivatives\footnote{%
Thanks to the relations between these geometrical objects, the other
building blocks, like e.g. the Riemann tensor and its descendants, are not
needed without any loss of generality.} is automatically invariant with
respect to (\ref{sGal_transform}). In order to construct the action, we need
also the invariant measures, namely, for the plus sign in (\ref%
{sGal_transform}), the measures $\mathrm{d}^{D}Z$ and $\mathrm{d}^{D}%
\overline{Z}$ 
\begin{equation}
\mathrm{d}^{D}Z=\mathrm{d}^{D}x\det \left( \eta +\frac{\mathrm{i}}{\alpha }%
\partial \partial \phi \right) ,~~~~\mathrm{d}^{D}\overline{Z}=\mathrm{d}%
^{D}x\det \left( \eta -\frac{\mathrm{i}}{\alpha }\partial \partial \phi
\right) ,  \label{dZdZbar}
\end{equation}%
or, for the minus sign in (\ref{sGal_transform}), the measures $\mathrm{d}%
^{D}Z^{+}$and $\mathrm{d}^{D}Z^{-}$ 
\begin{equation}
\mathrm{d}^{D}Z^{\pm }=\mathrm{d}^{D}x\det \left( \eta \pm \frac{1}{\alpha }%
\partial \partial \phi \right) ,  \label{dZ+dZ-}
\end{equation}%
and the canonical one%
\begin{equation}
\mathrm{d}^{D}x\sqrt{g}=\left\{ 
\begin{array}{c}
\sqrt{\mathrm{d}^{D}Z\mathrm{d}^{D}\overline{Z}} \\ 
\sqrt{\mathrm{d}^{D}Z^{+}\mathrm{d}^{D}Z^{-}}%
\end{array}%
\right. .  \label{canonical measure}
\end{equation}%
As discussed in \cite{Novotny:2016jkh}, the typical off-shell counterterm
action which is invariant with respect to the generalized polynomial shift
symmetry (\ref{sGal_transform}) is e.g. of the form%
\begin{equation}
S_{CT}^{\left( +\right) }=\int \sqrt{\mathrm{d}^{D}Z\mathrm{d}^{D}\overline{Z%
}}\mathcal{L}_{Z\overline{Z}}+\int \mathrm{d}^{D}Z\mathcal{L}_{Z}+\int 
\mathrm{d}^{D}\overline{Z}\mathcal{L}_{\overline{Z}}  \label{S_CT_plus}
\end{equation}%
where $\mathcal{L}_{Z}^{\ast }=\mathcal{L}_{\overline{Z}}$, or 
\begin{equation}
S_{CT}^{\left( -\right) }=\int \sqrt{\mathrm{d}^{D}Z^{+}\mathrm{d}^{D}Z^{-}}%
\mathcal{L}_{\pm }+\int \mathrm{d}^{D}Z^{+}\mathcal{L}_{+}+\int \mathrm{d}%
^{D}Z^{-}\mathcal{L}_{-}  \label{S_CT_minus}
\end{equation}%
according to the sign in (\ref{sGal_transform}) and (\ref{aGal metric}). The
functions $\mathcal{L}_{Z\overline{Z}}$, $\cdots ,\mathcal{L}_{-}$ are
diffeomorphism invariants built from the above geometrical building blocks.

However, this is not the whole story. As a consequence of the invariance of
the independent measures $\mathrm{d}^{D}Z^{+}$and $\mathrm{d}^{D}Z$ we can
costruct another invariant given as 
\begin{equation}
\sigma =\frac{\alpha }{2\mathrm{i}}\ln \left[ \frac{\det \left( \eta +\frac{%
\mathrm{i}}{\alpha }\partial \partial \phi \right) }{\det \left( \eta -\frac{%
\mathrm{i}}{\alpha }\partial \partial \phi \right) }\right]  \label{sigma+}
\end{equation}%
for the \textquotedblleft plus\textquotedblright\ branch of the
transformation (\ref{sGal_transform}) and 
\begin{equation}
\sigma =\frac{\alpha }{2}\ln \left[ \frac{\det \left( \eta +\frac{1}{\alpha }%
\partial \partial \phi \right) }{\det \left( \eta -\frac{1}{\alpha }\partial
\partial \phi \right) }\right]  \label{sigma-}
\end{equation}%
for the \textquotedblleft minus\textquotedblright\ branch. Note that these
invariants has been missed in \cite{Novotny:2016jkh}. The existence of these
invariants allows to forget the measures $\mathrm{d}^{D}Z$ and $\mathrm{d}%
^{D}\overline{Z}$ or $\mathrm{d}^{D}Z^{+}$and $\mathrm{d}^{D}Z^{-}$since
these can be constructed using the canonical measure (\ref{canonical measure}%
) and the functions of the invariants $\sigma $, e.g. 
\begin{equation}
\mathrm{d}^{D}Z=\mathrm{d}^{D}x\sqrt{g}\mathrm{e}^{\frac{\mathrm{i}}{\alpha }%
\sigma }.
\end{equation}%
The most general off shell counterterm action is then for the two branches
of the Special Galileon 
\begin{equation}
S_{CT}=\int \mathrm{d}^{D}x\sqrt{g}\mathcal{L}_{CT}\left( \sigma ,D\sigma
,\ldots ,\mathcal{K},D\mathcal{K},\ldots ,g^{\mu \nu }\right)
\label{general counterterm action}
\end{equation}%
where $\mathcal{L}_{CT}$ is diffeomorphism invariant built from the scalar $%
\sigma $, the extrinsic curvature tensor $\mathcal{K}_{\mu \nu \alpha }$,
their covariant derivatives, the inverse metric $g^{\mu \nu }$ and
Levi-Civita tensor $E^{\mu \nu \alpha \beta }$.

Remarkably, the transformation (\ref{aGal metric}) can be easily enlarged to
the case of additional non-Galileon fields $\psi _{\nu _{1}\ldots }^{\mu
_{1}\ldots }$. To get a minimal coupling of these fields with the Special
Galileon, one simply constructs the diffeomorphism invariant Lagrangians of
the non-Galileon fields on the general curved background and for the metric
then substitutes the effective metric (\ref{aGal metric}). The resulting
Lagrangians are then automatically invariant under the simultaneous
transformation (\ref{sGal_transform}) and the transformation of the $\psi
_{\nu _{1}\ldots }^{\mu _{1}\ldots }$ fields given schematically as%
\begin{equation}
\delta \psi _{\nu _{1}\ldots }^{\mu _{1}\ldots }=\theta \left[ -H^{\alpha
\beta }\partial _{\alpha }\phi \partial _{\beta }\psi _{\nu _{1}\ldots
}^{\mu _{1}\ldots }-\psi _{\alpha \ldots }^{\mu _{1}\ldots }H^{\alpha \beta
}\partial _{\beta }\partial _{\nu _{1}}\phi +\psi _{\nu _{1}\ldots }^{\alpha
\ldots }H^{\mu _{1}\beta }\partial _{\beta }\partial _{\alpha }\phi \ldots
+\ldots \right] ,  \label{non_Gal_transform}
\end{equation}%
and therefore can be added to the action given by the general formula (\ref%
{general counterterm action}). Note that the prescription (\ref%
{non_Gal_transform}) formally corresponds to the coordinate transformation%
\begin{equation}
\delta x^{\mu }=\theta H^{\mu \nu }\partial _{\nu }\phi .
\end{equation}%
For the fields, which carry also spinor indices, one needs additional
geometrical structures, namely the vielbain $m_{a}^{\mu }$ and the spinor
connection $\beta _{b\mu }^{a}$ which are expressed in terms of the second
and third derivatives of the Galileon field $\phi $ (see \cite%
{Novotny:2016jkh} for the explicit formulas). The invariance of the
resulting action with respect to the generalized polynomial shift symmetry (%
\ref{sGal_transform}) then guaranties under some additional assumptions%
\footnote{%
The suffucient condition is the absence of the cubic vertices.} the enhanced 
$O\left( p^{3}\right) $ soft Galileon limits of the tree level scattering
amplitudes.

Though the Special Galileon is well understood at the tree level,
considerably less is known about its true quantum properties. In the
literature, several one-loop calculations exists for the case of cubic
Galileon \cite%
{Brouzakis:2013lla,Brouzakis:2014bwa,Saltas:2016awg,Saltas:2016nkg,Goon:2016ihr}
and for general Galileon \cite{Kampf:2014rka, Heisenberg:2019udf}, but
systematic analysis with stress to the Special Galileon case is still
missing. Namely, it is not known, whether the loops do or do not break the
invariance with respect to the generalized polynomial shift symmetry (\ref%
{sGal_transform}) of the Special Galileon and whether the enhanced $O\left(
p^{3}\right) $ soft behavior survives the quantum corrections. In this paper
we initialize the studies in this direction and try to fill this gap
partially by means of explicit calculation of the UV divergent part of the
one-loop on-shell effective action using Dimensional Regularization (DR) and
providing the complete one-loop renormalization of the $S$-matrix. We also
calculate explicitly the lowest scattering amplitudes at one loop and
discuss their compatibility with other approaches.

The paper is organized as follows. In the section 2 we briefly remind the
basic facts concerning the functional approach to the $S-$matrix and its
relation to the on-shell effective action. Section 3 is devoted to the issue
of the quantum fluctuations in the classical Galileon background, we show
here that the fluctuations can be described with action which is manifestly
invariant with respect to the hidden Special Galileon symmetry. In the
section 4 we calculate the infinite part of the one-loop on-shell effective
action for the Special Galileon and give a complete classification of the
counterterms in $D=4$ for $i_{CT}^{\Gamma }\leq 6$. In the section 5 we
briefly discuss the extension of the soft theorem for Special Galileon to
one loop. The section 6 is devoted to explicit examples of the scattering
amplitudes. In section 7 we summarize and further discuss our results. The
technicalities concerning the classification of the counterterms are
presented in the appendix \ref{counterterms}

\section{The generating functional of the $S-$matrix and the on-shell
effective action}

For reader's convenience and in order to fix our notation let us first
briefly review the interrelation of the $S-$matrix and the on-shell
effective action (for original papers see \cite{Boulware:1968zz,
Arefeva:1974jv, Jevicki:1987ax}). The starting point for our calculation is
the general formula for the perturbative $S-$matrix, which we treat here as
the functional $\mathcal{S}\left[ \phi \right] $ of the external field $\phi 
$ and which is constructed as\footnote{%
Here and in what follows we often use the condesed notation where the dot
means integration over the corresponding spacetime coordinates, e.g. 
\begin{equation*}
\xi \cdot O\cdot \xi \equiv \int \mathrm{d}^{D}x\mathrm{d}^{D}y\xi \left(
x\right) O(x,y)\xi \left( y\right) .
\end{equation*}%
}%
\begin{equation}
\mathcal{S}\left[ \phi \right] =\exp \left( \frac{\mathrm{i}}{\hbar }%
\mathcal{T}\left[ \phi \right] \right) =\exp \left( -\frac{1}{2\hbar }\frac{%
\delta }{\delta \phi }\cdot \mathrm{i}\Delta _{F}\cdot \frac{\delta }{\delta
\phi }\right) \exp \left( \frac{\mathrm{i}}{\hbar }S_{\mathrm{int}}\left[
\phi \right] \right) .  \label{S-matrix}
\end{equation}%
Here $\mathrm{i}\Delta _{F}\left( x-y\right) $ is the Feynman propagator, $%
S_{int}\left[ \phi \right] $ is the interaction part of the action with UV
counterterms included and the functional differential operator%
\begin{equation}
\exp \left( -\frac{1}{2\hbar }\frac{\delta }{\delta \phi }\cdot \mathrm{i}%
\Delta _{F}\cdot \frac{\delta }{\delta \phi }\right) \equiv \exp \left( -%
\frac{1}{2\hbar }\int \mathrm{d}^{D}x\mathrm{d}^{D}y\frac{\delta }{\delta
\phi \left( x\right) }\mathrm{i}\Delta _{F}\left( x-y\right) \frac{\delta }{%
\delta \phi \left( y\right) }\right)
\end{equation}%
generates formally the chronological contractions of the perturbative Wick
expansion of the $S-$matrix. From the functional $\mathcal{S}\left[ \phi %
\right] $ we can derive the operator $S-$matrix in the Dirac interaction
picture inserting for $\phi $ the free field operators $\widehat{\phi }_{I}$
and treating all the operator product as normally ordered\footnote{%
For this reason, the functional \ $\mathcal{S}\left[ \phi \right] $ is
called the normal symbol of the $S-$matrix.}, namely%
\begin{equation}
\widehat{S}=\colon \mathcal{S}\left[ \widehat{\phi }_{I}\right] \colon
\end{equation}%
Therefore it is sufficient to know $\mathcal{S}\left[ \phi \right] $ on
shell, i.e. for the functional argument $\phi $ satisfying the free equation
of motion $\square \phi =0$.

The connected on-shell $n-$point scattering amplitudes $A_{n}\left(
p_{1},\ldots ,p_{n}\right) $ can be obtained directly from the functional $%
\mathcal{T}\left[ \phi \right] $ by means of the functional differetiatial
operation which is equivalent to the contraction of the operators $\widehat{%
\phi }_{I}$ with the creation or annihilation operators in the initial or
final state\footnote{%
For simplicity, in the following formula we tacitly assume all the particles
to be outgoing.}%
\begin{equation}
A_{n}\left( p_{1},\ldots ,p_{n}\right) =\int \prod\limits_{i=1}^{n}\mathrm{d}%
^{D}x_{i}\mathrm{e}^{\mathrm{i}p_{i}\cdot x_{i}}\frac{\delta }{\delta \phi
\left( x_{i}\right) }\mathcal{T}\left[ \phi \right] |_{\phi =0}.
\end{equation}%
Using Gaussian integration we can rewrite (\ref{S-matrix}) to the form which
formally includes the LSZ formulas%
\begin{equation}
\mathcal{S}\left[ \phi \right] =\exp \left( \frac{\mathrm{i}}{\hbar }S_{0}%
\left[ \phi \right] \right) \int D\varphi \exp \frac{\mathrm{i}}{\hbar }%
\left[ S\left[ \varphi \right] +\int \mathrm{d}^{D}x\varphi \overleftarrow{%
\square }\phi \right] ,
\end{equation}%
where $S_{0}\left[ \phi \right] =-\int \mathrm{d}^{D}x\phi \square \phi /2$
is the kinetic term and where $S\left[ \varphi \right] =S_{0}\left[ \varphi %
\right] +S_{\mathrm{int}}\left[ \varphi \right] $ is the complete action
with UV counterterms included 
\begin{equation}
S\left[ \varphi \right] =\int \mathrm{d}^{D}x\left( \mathcal{L}_{b}+\mathcal{%
L}_{CT}\right) \equiv S_{b}\left[ \varphi \right] +S_{CT}\left[ \varphi %
\right] =S_{0}\left[ \varphi \right] +S_{b,\mathrm{int}}\left[ \varphi %
\right] +S_{CT}\left[ \varphi \right]
\end{equation}%
In order to obtain the loop expansion of $\mathcal{T}\left[ \phi \right] $
let us substitute $\varphi =\phi _{cl}+\sqrt{\hbar }\xi $ where $\phi _{cl}$
is a solution of the integral equation 
\begin{equation}
\phi _{cl}=\phi +\frac{1}{\square }\frac{\delta S_{b,\mathrm{int}}\left[
\phi _{cl}\right] }{\delta \phi _{cl}},  \label{phi_cl definition}
\end{equation}%
and $\xi $ is a new integration variable of the functional integral.
Expanding now in powers of $\sqrt{\hbar }$ we get (up to an inessential
overall constant)%
\begin{eqnarray}
\mathcal{S}\left[ \phi \right] &=&\exp \frac{\mathrm{i}}{\hbar }\left( S_{0}%
\left[ \phi \right] +S\left[ \phi _{cl}\right] +\int \mathrm{d}^{D}x\phi
_{cl}\overleftarrow{\square }\phi \right)  \notag \\
&&\times \int D\xi \exp \left( \frac{\mathrm{i}}{2}\xi \cdot \frac{\delta
^{2}S_{b}\left[ \phi _{cl}\right] }{\delta \phi _{cl}\delta \phi _{c}}\cdot
\xi +O\left( \hbar ^{1/2}\right) \right)  \notag \\
&=&\exp \frac{\mathrm{i}}{\hbar }\left( S_{0}\left[ \phi \right] +S\left[
\phi _{cl}\right] +\int \mathrm{d}^{D}x\phi _{cl}\overleftarrow{\square }%
\phi \right)  \notag \\
&&\times \exp \left( \frac{\mathrm{i}}{2}\mathrm{Tr}\ln \frac{\delta
^{2}S_{b}\left[ \phi _{cl}\right] }{\delta \phi _{cl}\delta \phi _{c}}%
\right) \left( 1+O\left( \hbar \right) \right)
\end{eqnarray}%
Let us now expand the functional $\mathcal{T}\left[ \phi \right] $ and the
action $S\left[ \phi _{cl}\right] $ in powers $\hbar $ as 
\begin{eqnarray}
\mathcal{T}\left[ \phi \right] &=&\mathcal{T}^{\mathrm{tree}}\left[ \phi %
\right] +\hbar \mathcal{T}^{\mathrm{1-loop}}\left[ \phi \right] +O\left(
\hbar ^{2}\right) \\
S\left[ \phi _{cl}\right] &=&S_{b}\left[ \phi _{cl}\right] +\hbar S_{CT}^{%
\mathrm{1-loop}}\left[ \phi _{cl}\right] +O\left( \hbar ^{2}\right)
\end{eqnarray}%
where $S_{CT}^{\mathrm{1-loop}}=\int \mathrm{d}^{D}x\mathcal{L}_{CT}^{%
\mathrm{1-loop}}$ are the one-loop counterterms and $\mathcal{T}^{\mathrm{%
tree}}$ and $\mathcal{T}^{\mathrm{1-loop}}\left[ \phi \right] $ are the
tree- and one-loop level connected $S-$matrices respectively. We can then
identify 
\begin{eqnarray}
\mathcal{T}^{\mathrm{tree}}\left[ \phi \right] &=&S_{0}\left[ \phi \right]
+S_{b}\left[ \phi _{cl}\right] +\int \mathrm{d}^{D}x\phi _{cl}\overleftarrow{%
\square }\phi  \label{T^tree} \\
\mathcal{T}^{\mathrm{1-loop}}\left[ \phi \right] &=&\frac{\mathrm{i}}{2}%
\mathrm{Tr}\ln \frac{\delta ^{2}S_{b}\left[ \phi _{cl}\right] }{\delta \phi
_{cl}\delta \phi _{c}}+S_{CT}^{\mathrm{1-loop}}\left[ \phi _{cl}\right] .
\label{T^1-loop}
\end{eqnarray}%
Note that when understood as the functional of $\phi _{cl}$ the one-loop
connected $S-$matrix $\mathcal{T}^{\mathrm{1-loop}}\left[ \phi \right] $
coincides with the one-loop effective action $\Gamma ^{\mathrm{1-loop}}\left[
\phi _{cl}\right] $. Note also that for $\phi $ on shell (i.e. for $\square
\phi =0$) the field $\phi _{cl}$ satisfies the classical equation of motion
(EOM) 
\begin{equation}
\square \phi _{cl}=\frac{\delta S_{b,\mathrm{int}}\left[ \phi _{cl}\right] }{%
\delta \phi _{cl}}.  \label{EOM}
\end{equation}%
Therefore provided we are interested only in the on-shell scattering
amplitudes up to one loop order we need to know the one-loop on-shell
effective action $\Gamma ^{1-loop}\left[ \phi _{cl}\right] $. i.e. the one
loop effective action with argument satisfying the classical equation of
motion. This well known fact will be crucial for our following calculations.

\bigskip

\section{Quantum fluctuations in the classical background}

As discussed in the previous section, the key object for our further
calculation is the operator $\delta ^{2}S_{b}\left[ \phi _{cl}\right]
/\delta \phi _{cl}\delta \phi _{c}$ describing the propagation of quantum
fluctuation around a classical background $\phi _{cl}$ which satisfies the
classical equations of motion (\ref{EOM}). In the case of general Galileon
the equation of motion reads (cf. (\ref{lagrangian})) 
\begin{equation}
\frac{\delta S_{b}\left[ \phi _{cl}\right] }{\delta \phi _{cl}}%
=\sum_{n=0}^{D}\left( n+1\right) d_{n+1}\varepsilon ^{\mu _{1}\ldots \mu
_{D}}\varepsilon ^{\nu _{1}\ldots \nu _{D}}\prod_{i=1}^{n}\partial _{\mu
_{i}}\partial _{\nu _{i}}\phi _{cl}\prod_{j=n+1}^{D}\eta _{\mu _{j}\nu
_{j}}=0.  \label{EOM general case}
\end{equation}%
Taking the second functional derivative we derive for the fluctuation
operator 
\begin{eqnarray}
\frac{\delta ^{2}S_{b}\left[ \phi _{cl}\right] }{\delta \phi _{cl}\left(
x\right) \delta \phi _{cl}\left( y\right) } &=&\sum_{n=1}^{D}\left(
n+1\right) nd_{n+1}\varepsilon ^{\mu _{1}\ldots \mu _{D}}\varepsilon ^{\nu
_{1}\ldots \nu _{D}}\prod_{i=1}^{n-1}\partial _{\mu _{i}}\partial _{\nu
_{i}}\phi _{cl}\left( x\right)  \\
&&\times \prod_{j=n+1}^{D}\eta _{\mu _{j}\nu _{j}}\partial _{\mu
_{n}}\partial _{\nu _{n}}\delta ^{\left( D\right) }\left( x-y\right)   \notag
\\
&=&-G^{\mu \nu }\left[ \phi _{cl}\right] \partial _{\mu }\partial _{\nu
}\delta ^{\left( D\right) }\left( x-y\right) ,  \label{Gal_EOM}
\end{eqnarray}%
where we denoted%
\begin{eqnarray}
G^{\mu \nu }\left[ \phi _{cl}\right]  &=&-\frac{\partial }{\partial \left[
\partial _{\mu }\partial _{\nu }\phi _{cl}\right] }\frac{\delta S_{b}\left[
\phi _{cl}\right] }{\delta \phi _{cl}}  \notag \\
&=&\sum_{n=1}^{D}\left( n+1\right) nd_{n+1}\varepsilon ^{\mu _{1}\ldots \mu
_{D}}\varepsilon ^{\nu _{1}\ldots \nu _{D}}\prod_{i=1}^{n-1}\partial _{\mu
_{i}}\partial _{\nu _{i}}\phi _{cl}\left( x\right) \prod_{j=n+1}^{D}\eta
_{\mu _{j}\nu _{j}}.  \label{Gmn}
\end{eqnarray}%
Note, that $\partial _{\mu }G^{\mu \nu }\left[ \phi _{cl}\right] =\partial
_{\nu }G^{\mu \nu }\left[ \phi _{cl}\right] =0$, and therefore the quantum
fluctuations in the quadratic approximation are described by the action%
\begin{eqnarray}
S_{q}\left[ \phi _{cl},\xi \right]  &=&\frac{1}{2}\xi \cdot \frac{\delta
^{2}S_{b}\left[ \phi _{cl}\right] }{\delta \phi _{cl}\delta \phi _{c}}\cdot
\xi   \notag \\
&=&\frac{1}{2}\int \mathrm{d}^{D}xG^{\mu \nu }\left[ \phi _{cl}\right]
\left( x\right) \partial _{\mu }\xi \left( x\right) \partial _{\nu }\xi
\left( x\right) ,  \label{fluctuation action}
\end{eqnarray}

In the case of Special Galileon we can sum up the right hand side of (\ref%
{Gal_EOM}) and (\ref{Gmn}) in a closed form. Indeed, inserting (\ref{sGal1})
with $\beta =0$ we get for the first branch of the Galileon Lagrangians%
\footnote{%
We have rescaled the coefficints $d_{n}^{\left( \pm \right) }$ by a factor $%
\pm \left( -1\right) ^{D}/D!$ in comparisson with (\ref{sGal1}) and (\ref%
{sGal2}) in order to ensure the canonical normalization of the kinetic term.}%
\begin{eqnarray}
\frac{\delta S_{b}^{\left( +\right) }\left[ \phi _{cl}\right] }{\delta \phi
_{cl}} &=&-\frac{\mathrm{i}\alpha \left( -1\right) ^{D}}{D!}\sum_{n=1}^{%
\left[ \frac{D}{2}\right] }\left( 
\begin{array}{c}
D \\ 
2n-1%
\end{array}%
\right) \left( \frac{\mathrm{i}}{\alpha }\right) ^{2n-1}\varepsilon ^{\mu
_{1}\ldots \mu _{D}}\varepsilon ^{\nu _{1}\ldots \nu
_{D}}\prod_{i=1}^{2n-1}\partial _{\mu _{i}}\partial _{\nu _{i}}\phi
_{cl}\prod_{j=2n-1}^{D}\eta _{\mu _{j}\nu _{j}}  \notag \\
&=&-\frac{\mathrm{i}\alpha \left( -1\right) ^{D}}{2D!}\varepsilon ^{\mu
_{1}\ldots \mu _{D}}\varepsilon ^{\nu _{1}\ldots \nu _{D}}\left[
\prod_{i=1}^{D}\left( \eta _{\mu _{i}\nu _{i}}+\frac{\mathrm{i}}{\alpha }%
\partial _{\mu _{i}}\partial _{\nu _{i}}\phi _{cl}\right)
-\prod_{i=1}^{D}\left( \eta _{\mu _{i}\nu _{i}}-\frac{\mathrm{i}}{\alpha }%
\partial _{\mu _{i}}\partial _{\nu _{i}}\phi _{cl}\right) \right]   \notag
\end{eqnarray}%
Using the general identity valid for any $D\times D$ matrix $M_{\mu \nu }$%
\begin{equation}
\varepsilon ^{\mu _{1}\ldots \mu _{D}}\varepsilon ^{\nu _{1}\ldots \nu
_{D}}\prod_{i=1}^{D}M_{\mu _{i}\nu _{i}}=D!\det M_{\mu \nu }=\left(
-1\right) ^{D-1}D!\det M_{\nu }^{\mu },
\end{equation}%
we can rewrite the EOM in a compact form as%
\begin{equation}
\frac{\delta S_{b}^{\left( +\right) }\left[ \phi _{cl}\right] }{\delta \phi
_{cl}}=\frac{\mathrm{i}\alpha }{2}\left( \mathcal{D}_{+}^{\left( +\right) }-%
\mathcal{D}_{-}^{\left( +\right) }\right) =0,  \label{EOM_+}
\end{equation}%
where we denoted%
\begin{equation}
\mathcal{D}_{\pm }^{\left( +\right) }=\det \left( \delta _{\nu }^{\mu }\pm 
\frac{\mathrm{i}}{\alpha }\partial ^{\mu }\partial _{\nu }\phi _{cl}\right) .
\label{Det_plus}
\end{equation}%
In the same way, using instead (\ref{sGal2}) for the second branch of the
Special Galileon Lagrangians, we get%
\begin{equation}
\frac{\delta S_{b}^{\left( -\right) }\left[ \phi _{cl}\right] }{\delta \phi
_{cl}}=-\frac{\alpha }{2}\left( \mathcal{D}_{+}^{\left( -\right) }-\mathcal{D%
}_{-}^{\left( -\right) }\right) =0,  \label{EOM_-}
\end{equation}%
where now%
\begin{equation}
\mathcal{D}_{\pm }^{\left( -\right) }=\det \left( \delta _{\nu }^{\mu }\pm 
\frac{1}{\alpha }\partial ^{\mu }\partial _{\nu }\phi _{cl}\right) .
\end{equation}%
Taking the partial derivative of (\ref{EOM_+}) and (\ref{EOM_-}) with
respect to $\partial _{\mu }\partial _{\nu }\phi _{cl}$ we get%
\begin{eqnarray}
G_{\left( +\right) }^{\mu \nu }\left[ \phi _{cl}\right]  &=&\frac{1}{2}\eta
^{\mu \alpha }\left[ \mathcal{D}_{+}^{\left( +\right) }\left( \boldsymbol{1}+%
\frac{\mathrm{i}}{\alpha }\boldsymbol{\partial \partial \phi }_{cl}\right)
^{-1}+\mathcal{D}_{-}^{\left( +\right) }\left( \boldsymbol{1}-\frac{\mathrm{i%
}}{\alpha }\boldsymbol{\partial \partial \phi }_{cl}\right) ^{-1}\right]
_{\alpha }^{\nu } \\
G_{\left( -\right) }^{\mu \nu }\left[ \phi _{cl}\right]  &=&\frac{1}{2}\eta
^{\mu \alpha }\left[ \mathcal{D}_{+}^{\left( -\right) }\left( \boldsymbol{1}+%
\frac{1}{\alpha }\boldsymbol{\partial \partial \phi }_{cl}\right) ^{-1}+%
\mathcal{D}_{-}^{\left( -\right) }\left( \boldsymbol{1}-\frac{1}{\alpha }%
\boldsymbol{\partial \partial \phi }_{cl}\right) ^{-1}\right] _{\alpha
}^{\nu }
\end{eqnarray}%
where $\boldsymbol{\partial \partial \phi }_{cl}$ denotes the $D\times D$
matrix with elements $\partial ^{\mu }\partial _{\nu }\phi _{cl}$. \ This
can be further simplified using the EOM (\ref{EOM_+}) and (\ref{EOM_-}) in
the form\footnote{%
Note, that EOM ensures that $\mathcal{D}_{+}^{\left( +\right) }=\mathcal{D}%
_{-}^{\left( +\right) }$ is real. Here we further suppose that $\mathcal{D}%
_{+}^{\left( +\right) }$ is positive as suggests the weak field limit $%
\mathcal{D}_{+}^{\left( +\right) }=1+O\left( \partial \partial \phi
_{cl}\right) $} 
\begin{equation}
\mathcal{D}_{+}^{\left( +\right) }=\mathcal{D}_{-}^{\left( +\right) }=\sqrt{%
\mathcal{D}_{+}^{\left( +\right) }\mathcal{D}_{-}^{\left( +\right) }}
\label{EOM for determinants}
\end{equation}%
with the result%
\begin{eqnarray}
G_{\left( +\right) }^{\mu \nu }\left[ \phi _{cl}\right]  &=&\eta ^{\mu
\alpha }\sqrt{\mathcal{D}_{+}^{\left( +\right) }\mathcal{D}_{-}^{\left(
+\right) }}\left[ \left( \boldsymbol{1}+\frac{1}{\alpha ^{2}}\boldsymbol{%
\partial \partial \phi }_{cl}\cdot \boldsymbol{\partial \partial \phi }%
_{cl}\right) ^{-1}\right] _{\alpha }^{\nu }  \notag \\
&=&\sqrt{g}g^{\mu \nu },  \label{Gmn in terms of gmn}
\end{eqnarray}%
and similarly $G_{\left( -\right) }^{\mu \nu }\left[ \phi _{cl}\right] =%
\sqrt{g}g^{\mu \nu }$. Here $g^{\mu \nu }$is the matrix inverse of the
Special Galileon effective metric $g_{\mu \nu }=\eta _{\mu \nu }\pm \partial
_{\mu }\partial \phi _{cl}\cdot $ $\partial \partial _{\nu }\phi
_{cl}/\alpha ^{2}$ (see (\ref{aGal metric})) and $g=\left\vert \det g_{\mu
\nu }\right\vert $ is its determinant. The fluctuation action (\ref%
{fluctuation action}) is thus rewritten in the geometrical form\footnote{%
Let us note, that in the case of general Galileon in $D\neq 2$ dimensions,
the fluctuation action (\ref{fluctuation action}) can be also rewritten in
the geometric form 
\begin{equation*}
S_{q}\left[ \phi _{cl},\xi \right] =\frac{1}{2}\int \mathrm{d}^{D}x\sqrt{%
\mathcal{G}}\mathcal{G}^{\mu \nu }\partial _{\mu }\xi \partial _{\nu }\xi 
\end{equation*}%
where now 
\begin{equation*}
\mathcal{G}^{\mu \nu }=\left\vert \det G^{\mu \nu }\left[ \phi _{cl}\right]
\right\vert ^{\frac{1}{2-D}}G^{\mu \nu }\left[ \phi _{cl}\right] .
\end{equation*}%
The exclusivity of the Special Galileons lies in the fact, that in this case
the inverse metric $G_{\mu \nu }$ is computable and equals to the explicitly
known effective metric $g_{\mu \nu }$. This ensures the invariance of the
fluctuation action with respect to the hidden Special Galileon symmetry.}%
\begin{equation}
S_{q}\left[ \phi _{cl},\xi \right] =\frac{1}{2}\int \mathrm{d}^{D}x\sqrt{g}%
g^{\mu \nu }\partial _{\mu }\xi \partial _{\nu }\xi .
\label{Geometric action for fluctuations}
\end{equation}%
Therefore it is invariant under the hidden Special Galileon transformation (%
\ref{sGal_transform}) of the Galileon field $\phi _{cl}$ and simultaneous
transformation of the field $\xi $ according to (cf. (\ref{non_Gal_transform}%
))%
\begin{equation}
\delta \xi =-\theta H^{\alpha \beta }\partial _{\alpha }\phi _{cl}\partial
_{\beta }\xi .  \label{xi transformation}
\end{equation}%
This can be used for a formal proof of the invariance of the pure loop part
of the on-shell effective action $\Gamma _{\mathrm{loop}}^{\mathrm{1-loop}}%
\left[ \phi _{cl}\right] $ with respect to the hidden Special Galileon
symmetry. Indeed, up to an inessential constant\footnote{%
From now on we return to the natural units $\hbar =1$.}, 
\begin{equation}
\Gamma _{\mathrm{loop}}^{\mathrm{1-loop}}\left[ \phi _{cl}\right] =-\mathrm{i%
}\ln \int D\xi \exp \left( \mathrm{i}S_{q}\left[ \phi _{cl},\xi \right]
\right) 
\end{equation}%
where appropriate regularization compatible with the symmetry (e.g. the
dimensional regularization will do the job as we will discussed later) is
tacitly assumed. Thus under the transformations (\ref{sGal_transform}) and (%
\ref{xi transformation})%
\begin{eqnarray}
\Gamma _{\mathrm{loop}}^{\mathrm{1-loop}}\left[ \phi _{cl}+\delta \phi _{cl}%
\right]  &=&-\mathrm{i}\ln \int D\left( \xi +\delta \xi \right) \exp \left( 
\mathrm{i}S_{q}\left[ \phi _{cl}+\delta \phi _{cl},\xi +\delta \xi \right]
\right)   \label{Gamma_invariance} \\
&=&-\mathrm{i}\ln \int D\xi J\exp \left( \mathrm{i}S_{q}\left[ \phi
_{cl},\xi \right] \right) 
\end{eqnarray}%
where the Jacobian $J=1-\theta \mathrm{Tr}\left( H^{\alpha \beta }\partial
_{\alpha }\phi _{cl}\partial _{\beta }\right) $ can be shown to be equal to
one (this holds within the dimensional regularization - see e.g. similar
calculation in \cite{Kampf:2014rka}), which proves the statement.

In what follows we will concentrate on the UV divergent part of $\Gamma _{%
\mathrm{loop}}^{\mathrm{1-loop}}\left[ \phi _{cl}\right] _{\infty }$ which
is local and which determines the UV\ divergent part of the one-loop
counterterm action $S_{CT}^{\mathrm{1-loop}}\left[ \phi _{cl}\right] $%
\begin{equation*}
S_{CT}^{\mathrm{1-loop}}\left[ \phi _{cl}\right] =-\Gamma _{\mathrm{loop}}^{%
\mathrm{1-loop}}\left[ \phi _{cl}\right] _{\infty }+S_{CT}^{\mathrm{1-loop}}%
\left[ \phi _{cl}\right] _{\mathrm{finite}}
\end{equation*}%
where $S_{CT}^{\mathrm{1-loop}}\left[ \phi _{cl}\right] _{\mathrm{finite}}$
includes all the finite parts of the couterterms. According to (\ref%
{Gamma_invariance}), the divergent part of $S_{CT}^{\mathrm{1-loop}}\left[
\phi _{cl}\right] $ is invariant with respect to the hidden Special Galileon
symmetry. In the next section we prove this statement less formally by
explicit calculation of $\Gamma _{\mathrm{loop}}^{\mathrm{1-loop}}\left[
\phi _{cl}\right] _{\infty }$ within dimensional regularization.

\bigskip

\section{The UV divergences at one loop\label{UV divergences at one loop}}

The most economic way how to obtain the UV divergent part of $\Gamma _{%
\mathrm{loop}}^{\mathrm{1-loop}}\left[ \phi _{cl}\right] $ within the
dimensional regularization is to use the master formula stemming from the
heat kernel expansion \cite{tHooft:1973bhk, Lee:1984ud}. From (\ref%
{Geometric action for fluctuations}) it follows that 
\begin{equation}
S_{q}\left[ \phi _{cl},\xi \right] =-\frac{1}{2}\int \mathrm{d}^{D}x\sqrt{g}%
\xi g^{\mu \nu }D_{\mu }D_{\nu }\xi
\end{equation}%
where $D_{\mu }$ is the covariant derivative associated with the effective
metric $g_{\mu \nu }$. Therefore, up to an inessential constant, we can
express $\Gamma _{\mathrm{loop}}^{\mathrm{1-loop}}\left[ \phi _{cl}\right] $
in terms of the heat kernel (here we use the Lorentzian signature, see e.g. 
\cite{DeWitt:2003pm}) 
\begin{equation}
K\left( \tau ;x,y\right) =\exp \left[ \mathrm{i}\tau \left( g^{\mu \nu
}D_{\mu }D_{\nu }+\mathrm{i}0\right) \right] ,
\end{equation}%
using the well known formula for logarithm of the operator determinant 
\begin{equation}
\Gamma _{\mathrm{loop}}^{\mathrm{1-loop}}\left[ \phi _{cl}\right] =\frac{%
\mathrm{i}}{2}\int \mathrm{d}^{D}x\sqrt{g}\int_{0}^{\infty }\frac{\mathrm{d}%
\tau }{\tau }K\left( \tau ;x,x\right) .  \label{heat kernel representation}
\end{equation}%
At the coincident points $x=y$ the heat kernel $K\left( \tau ;x,y\right) $
has the following asymptotic expansion for $\tau \rightarrow 0$%
\begin{equation}
K\left( \tau ;x,x\right) =\frac{\mathrm{i}}{\left( 4\pi \mathrm{i}\tau
\right) ^{D/2}}\sum\limits_{n=0}^{\infty }\left( \mathrm{i}\tau \right)
^{n}a_{n}\left( x\right) \mathrm{e}^{-\tau 0}.
\end{equation}%
where $a_{n}\left( x\right) $'s are the coincidence limits of the
Seeley-DeWitt coefficients. The UV divergences of the effective action $%
\Gamma _{loop}^{1-loop}\left[ \phi _{cl}\right] $ are then connected with
the lower limit of the integral in the formula (\ref{heat kernel
representation}) and can be cured by dimensional regularization writing $%
D\rightarrow D-2\varepsilon $. This gives the master formula valid for $D$
even\footnote{%
For $D$ odd all the one loop infinities are removed automatically by means
of the dimensional continuation.} 
\begin{equation}
\Gamma _{\mathrm{loop}}^{\mathrm{1-loop}}\left[ \phi _{cl}\right] _{\infty }=%
\frac{\mu ^{-2\varepsilon }}{2\left( 4\pi \right) ^{D/2}}\frac{1}{%
\varepsilon }\int \mathrm{d}^{D}x\sqrt{g}\,a_{D/2}\left( x\right) .
\label{master formula}
\end{equation}%
where $\mu $ is the dimensional renormalization scale parameter. The
Seeley-DeWitt coefficients $a_{n}\left( x\right) $ are calculable and
explicitly known for $n\leq 10$ in terms of the geometrical invariants (for
a review and a complete list of references see \cite{Vassilevich:2003xt}),
e.g. 
\begin{eqnarray}
a_{0} &=&1  \notag \\
a_{1} &=&\frac{1}{6}R  \notag \\
a_{2} &=&\frac{1}{180}\left( R_{\alpha \beta \mu \nu }R^{\alpha \beta \mu
\nu }-R_{\mu \nu }R^{\mu \nu }\right) +\frac{1}{72}R^{2}+\frac{1}{30}g^{\mu
\nu }D_{\mu }D_{\nu }R  \notag \\
a_{3} &=&\frac{1}{7!}\left( 18g^{\mu \nu }g^{\alpha \beta }D_{\mu }D_{\nu
}D_{\alpha }D_{\beta }R+17g^{\mu \nu }D_{\mu }RD_{\nu }R+\ldots \right) .
\label{Seely-DeWitt}
\end{eqnarray}%
Here $R_{\alpha \beta \mu \nu }$, $R_{\mu \nu }$and $R$ are the Riemann
curvature tensor, the Ricci tensor and the scalar curvature corresponding in
our case to the effective metric $g_{\mu \nu }$. All these objects can be
expressed in terms of the extrinsic curvature tensor (\ref{extrinsic
curvature}) and the inverse metric $g^{\mu \nu }$, namely (see \cite%
{Novotny:2016jkh}\ for more details)%
\begin{equation}
R_{\alpha \beta \mu \nu }=g^{\rho \sigma }\left( \mathcal{K}_{\rho \mu
\alpha }\mathcal{K}_{\sigma \nu \beta }-\mathcal{K}_{\rho \mu \beta }%
\mathcal{K}_{\sigma \nu \alpha }\right) .  \label{Riemann tensor}
\end{equation}%
As discussed in the introduction, due to their geometrical nature, all the
Seeley-DeWitt coefficients are then automatically invariant with respect to
the hidden Special Galileon symmetry and so is the divergent part of the
one-loop on shell effective action (\ref{master formula}).

In what follows we will concentrate on the four-dimensional case. Note that
we can further simplify the above formulas dropping the last term in $%
a_{2}\left( x\right) $ which is a total derivative and also eliminating a
total derivative corresponding to the four dimensional Gauss-Bonett term%
\begin{equation}
G=R_{\alpha \beta \mu \nu }R^{\alpha \beta \mu \nu }-4R_{\mu \nu }R^{\mu \nu
}+R^{2}.  \label{gauss bonnet}
\end{equation}%
The result then reads%
\begin{equation}
\Gamma _{\mathrm{loop}}^{\mathrm{1-loop}}\left[ \phi _{cl}\right] _{\infty
}^{D=4}=\frac{\mu ^{-2\varepsilon }}{\left( 4\pi \right) ^{2}}\frac{1}{120}%
\frac{1}{\varepsilon }\int \mathrm{d}^{4}x\sqrt{g}\left( R_{\mu \nu }R^{\mu
\nu }+\frac{1}{2}R^{2}\right) .  \label{4D divergences}
\end{equation}%
Note that the index of the vertices of (\ref{4D divergences}) equals to
(note that $g_{\mu \nu }$ contain exactly two derivatives per field while $%
\mathcal{K}_{\sigma \nu \beta }$ has three derivatives per field and that $%
RR\sim \mathcal{K}^{4}$) 
\begin{equation*}
i_{V}=\left( 2E_{V}+4\right) -\left( 2E_{V}-2\right) =6.
\end{equation*}%
This coincides with the index of the corresponding one-loop graphs (cf (\ref%
{index of graph})). The dimensional regularization therefore respects the
hierarchy of the counterterms, especially the basic Lagrangian is not
renormalized. This is not true for other regularization schemes, e.g. for
the momentum cutoff corresponding to the deformation of the integral in (\ref%
{heat kernel representation}) by means of introducing lower limit of the
integration of the order $1/\Lambda ^{2}$. In such a case all the
Seeley-DeWitt coefficients up to $n=2$ (or $n=D/2$ for general $D$)
contribute to $\Gamma _{\mathrm{loop}}^{\mathrm{1-loop}}\left[ \phi _{cl}%
\right] _{\infty }$.

Because we are concentrated on the $S-$matrix and on shell amplitudes, we
can simplify (\ref{4D divergences}) further using the classical EOM. Indeed,
taking a derivative of (\ref{EOM general case}), we get%
\begin{equation}
\sum_{n=0}^{D}\left( n+1\right) nd_{n+1}\varepsilon ^{\mu _{1}\ldots \mu
_{D}}\varepsilon ^{\nu _{1}\ldots \nu _{D}}\prod_{i=1}^{n-1}\partial _{\mu
_{i}}\partial _{\nu _{i}}\phi _{cl}\prod_{j=n+1}^{D}\eta _{\mu _{j}\nu
_{j}}\partial _{\alpha }\partial _{\mu _{n}}\partial _{\nu _{n}}\phi _{cl}=0
\end{equation}%
which can be rewritten with help of (\ref{Gmn}), (\ref{Gmn in terms of gmn})
and (\ref{extrinsic curvature}) as 
\begin{equation}
-\alpha \sqrt{g}g^{\mu \nu }\mathcal{K}_{\mu \nu \alpha }=0.
\label{EOM for K}
\end{equation}%
The Ricci tensor given by eq. (\ref{Riemann tensor}) then simplifies 
\begin{equation}
R_{\mu \nu }=-g^{\alpha \beta }g^{\rho \sigma }\mathcal{K}_{\rho \beta \mu }%
\mathcal{K}_{\sigma \nu \alpha }\equiv -\frac{1}{\alpha ^{2}}\partial _{\mu
}\partial \partial \phi _{cl}\boldsymbol{\colon }\partial \partial \partial
_{\nu }\phi _{cl}  \label{Ricci tensor}
\end{equation}%
where the bold colon (and in general the bold dot in what follows) denotes
the contraction of the adjacent indices with help of the inverse effective
metric $g^{\kappa \lambda }$. Using the same shorthand notation we get for
the scalar curvature%
\begin{equation}
R=-\langle \langle \mathcal{K}\boldsymbol{\colon }\mathcal{K}\rangle \rangle
=-\frac{1}{\alpha ^{2}}\langle \langle \partial \partial \partial \phi _{cl}%
\boldsymbol{\colon }\partial \partial \partial \phi _{cl}\rangle \rangle
\label{Ricci scalar}
\end{equation}%
where $\langle \langle \cdot \rangle \rangle $ denotes the trace with
respect to the inverse metric $g^{\kappa \lambda }$. Finally we get 
\begin{equation}
\Gamma _{\mathrm{loop}}^{\mathrm{1-loop}}\left[ \phi _{cl}\right] _{\infty
}^{D=4}=\frac{\mu ^{-2\varepsilon }}{\left( 4\pi \right) ^{2}}\frac{1}{120}%
\frac{1}{\varepsilon }\int \mathrm{d}^{4}x\sqrt{g}\left[ \langle \langle 
\mathcal{K}\boldsymbol{\colon }\mathcal{K}\boldsymbol{\cdot }\mathcal{K}%
\boldsymbol{\colon }\mathcal{K}\rangle \rangle +\frac{1}{2}\langle \langle 
\mathcal{K}\boldsymbol{\colon }\mathcal{K}\rangle \rangle ^{2}\right] ,
\end{equation}%
or more explicitly%
\begin{eqnarray}
\Gamma _{\mathrm{loop}}^{\mathrm{1-loop}}\left[ \phi _{cl}\right] _{\infty
}^{D=4} &=&\frac{\mu ^{-2\varepsilon }}{\left( 4\pi \right) ^{2}}\frac{1}{%
120\alpha ^{4}}\frac{1}{\varepsilon }\int \mathrm{d}^{4}x\sqrt{\left\vert
\det \left( \eta \pm \frac{1}{\alpha ^{2}}\partial \partial \phi _{cl}%
\boldsymbol{\cdot }\partial \partial \phi _{cl}\right) \right\vert }  \notag
\\
&&\times \left[ \langle \langle \partial \partial \partial \phi _{cl}%
\boldsymbol{\colon }\partial \partial \partial \phi _{cl}\boldsymbol{\cdot }%
\partial \partial \partial \phi _{cl}\boldsymbol{\colon }\partial \partial
\partial \phi _{cl}\rangle \rangle \right.  \notag \\
&&\left. +\frac{1}{2}\langle \langle \partial \partial \partial \phi _{cl}%
\boldsymbol{\colon }\partial \partial \partial \phi _{cl}\rangle \rangle ^{2}%
\right] .
\end{eqnarray}%
The two operators on the right hand sides of the last two equations are
therefore the only operators invariant with respect to the hidden Special
Galileon symmetry which have to be inserted with infinite coefficients into
the counterterm action $S_{CT}^{\mathrm{1-loop}}$. Interestingly, there are
no other non-vanishing invariant operators of the form ${\mathcal{K}}^{4}$
provided the EOM constraint (\ref{EOM for K}) is satisfied\footnote{%
See appendix \ref{counterterms} for more details.}.

The one-loop counterterm action which is necessary to renormalize the
infinities of the on-shell amplitudes up to and including the graphs with
the graph index $i_{CT}^{\Gamma }\leq 6$ (cf. (\ref{index of graph}))
therefore reads%
\begin{equation}
S_{CT}^{\mathrm{1-loop}}=\frac{\mu ^{-2\varepsilon }}{\left( 4\pi \right)
^{2}}\frac{1}{120}\frac{\alpha ^{2}}{\Lambda ^{6}}\int \mathrm{d}^{4}x\sqrt{g%
}\left[ c_{1}\mathcal{O}_{1}+c_{2}\mathcal{O}_{2}\right]
+S_{CT,~~finite}^{1-loop},  \label{CT infinite}
\end{equation}%
where we have denoted 
\begin{eqnarray}
\mathcal{O}_{1} &=&R_{\mu \nu }R^{\mu \nu }=\langle \langle \mathcal{K}%
\boldsymbol{\colon }\mathcal{K}\boldsymbol{\cdot }\mathcal{K}\boldsymbol{%
\colon }\mathcal{K}\rangle \rangle  \label{O1} \\
\mathcal{O}_{2} &=&R^{2}=\langle \langle \mathcal{K}\boldsymbol{\colon }%
\mathcal{K}\rangle \rangle ^{2}  \label{O2}
\end{eqnarray}%
In the formula (\ref{CT infinite}), $\Lambda $ is the scale which controls
the systematic derivative expansion, or more precisely the expansion in the
graph index $i_{CT}^{\Gamma }$, and $c_{i}$ are dimensionless bare couplings
to be specified later. The finite part of the counterterm action $S_{CT,~~%
\mathrm{finite}}^{\mathrm{1-loop}}$ has the general form (\ref{general
counterterm action}) . However, provided the EOM (\ref{EOM_+}) or (\ref%
{EOM_-}) are satisfied, the invariant $\sigma $ and their covariant
derivatives do not correspond to the independent building blocs. Indeed,
using the EOM in the form (\ref{EOM_+}) we get from (\ref{sigma+}) for the
\textquotedblleft plus\textquotedblright\ branch%
\begin{equation}
\sigma =\frac{\alpha }{2\mathrm{i}}\ln \left[ \frac{\mathcal{D}_{+}^{\left(
+\right) }}{\mathcal{D}_{-}^{\left( +\right) }}\right] =0.
\end{equation}%
and analogicaly for the \textquotedblleft minus\textquotedblright\ branch.
As a result, $S_{CT,~~\mathrm{finite}}^{\mathrm{1-loop}}$ can be written
without loss of generality in the form 
\begin{equation}
S_{CT,~~\mathrm{finite}}^{\mathrm{1-loop}}\mathcal{=}\mu ^{-2\varepsilon
}\int \mathrm{d}^{4}x\sqrt{g}\sum_{j>2}\frac{\alpha ^{2}}{\Lambda ^{i_{%
\mathcal{O}_{j}}}}c_{j}\mathcal{O}_{j},  \label{CT finite}
\end{equation}%
where the operators $\mathcal{O}_{j}$, $j>2$ are given schematically as $%
\mathcal{O}_{j}=D^{n_{j}}\mathcal{K}^{m_{j}}$. The index of the
corresponding terms in the action is then simply\footnote{%
Note that the measure $\mathrm{d}^{4}x\sqrt{g}$ does not contribute to the
index since it contains two derivatives per field.}%
\begin{equation}
i_{\mathcal{O}_{j}}=n_{j}+m_{j}+2
\end{equation}%
The operators $\mathcal{O}_{j}$, $j>2$ should form together with $\mathcal{O}%
_{1}$ and $\mathcal{O}_{2}$ a complete\footnote{%
Here complete means modulo integration by parts and use of EOM.} set of
operators with index $i_{\mathcal{O}_{j}}\leq 6$ which are invariant with
respect to the hidden Special Galileon symmetry. There are only two
operators with $i_{\mathcal{O}_{j}}<6$ which do not vanish as a consequence
of EOM (\ref{EOM for K}), namely\footnote{%
These two operators correspond to the quartic and quadratic divergences
respectively when the momentum cutoff is used instead of dimensional
regularization.} 
\begin{eqnarray}
\mathcal{O}_{3} &=&\langle \langle \mathcal{K}\boldsymbol{\colon }\mathcal{K}%
\rangle \rangle =-R, \\
\mathcal{O}_{4} &=&1,
\end{eqnarray}%
with $i_{\mathcal{O}_{3}}=4$ and $i_{\mathcal{O}_{4}}=2$ respectively (cf. 
\cite{Novotny:2016jkh} for further discussion). For $i_{\mathcal{O}_{j}}=6$
there is on top of $\mathcal{O}_{1}$ and $\mathcal{O}_{2}$ only one (up to
integration by parts) invariant operator with one covariant derivative%
\begin{equation}
\mathcal{O}_{5}=\langle \langle \left( \mathcal{K}\boldsymbol{\colon }%
\mathcal{K}\right) \boldsymbol{\cdot }\left( D\boldsymbol{\cdot }\mathcal{K}%
\right) \mathcal{\rangle \rangle },
\end{equation}%
and we get two invariant operators with two covariant derivatives%
\begin{eqnarray}
\mathcal{O}_{6} &=&\left( D\boldsymbol{\cdot }\mathcal{K}\right) \boldsymbol{%
\colon }\left( D\boldsymbol{\cdot }\mathcal{K}\right) \\
\mathcal{O}_{7} &=&\langle \langle \left( \mathcal{K}\boldsymbol{\colon }%
\left( D\boldsymbol{\cdot }D\right) \mathcal{K}\right) \mathcal{\rangle
\rangle }.
\end{eqnarray}%
However, as shown in the appendix \ref{counterterms}, the operators $%
\mathcal{O}_{5}$ and $\mathcal{O}_{6}$ vanish due to the relation%
\begin{equation}
D\boldsymbol{\cdot }\mathcal{K}_{\mu \nu }=g^{\alpha \beta }D_{\alpha }%
\mathcal{K}_{\beta \mu \nu }=0,
\end{equation}%
which is valid as a consequence of EOM, and the operator $\mathcal{O}_{7}$
can be rewritten (again using EOM and integration by parts) as a linear
combination of the operators $\mathcal{O}_{1}$ and $\mathcal{O}_{2}$ plus a
total derivative. The same is true for the apparently independent operator 
\begin{equation}
\mathcal{O}_{8}=g^{\kappa \rho }g^{\gamma \beta }g^{\delta \alpha }g^{\mu
\nu }\mathcal{K}_{\kappa \gamma \delta }\mathcal{K}_{\nu \rho }\boldsymbol{%
\cdot }\mathcal{K}_{\beta }\boldsymbol{\cdot }\mathcal{K}_{\alpha \mu }.
\end{equation}%
The parity odd operators with $i_{\mathcal{O}_{j}}\leq 6$ of the type%
\begin{equation}
\mathcal{O}_{j}=\left( D^{n_{j}}\mathcal{K}^{m_{j}}\right) _{\mu _{1}\mu
_{2}\mu _{3}\mu _{4}}E^{\mu _{1}\mu _{2}\mu _{3}\mu _{4}},
\label{parity odd}
\end{equation}%
where $E^{\mu _{1}\mu _{2}\mu _{3}\mu _{4}}$ is the contravariant
Levi-Civita tensor (\ref{levi-civita}), vanish on-shell due to the symmetry
properties of the building blocks (see appendix \ref{counterterms}).

To summarize, there are only four independent on-shell counterterms,
interestingly just those which already appeared in the Seeley-DeWitt
coefficients (\ref{Seely-DeWitt})%
\begin{eqnarray}
S_{CT}^{\mathrm{1-loop}} &\mathcal{=}&\mu ^{-2\varepsilon }\int \mathrm{d}%
^{4}x\sqrt{g}\sum_{j=1}^{4}\frac{\alpha ^{2}}{\Lambda ^{i_{\mathcal{O}_{j}}}}%
c_{j}\mathcal{O}_{j}  \notag \\
&=&\mu ^{-2\varepsilon }\frac{\alpha ^{2}}{\Lambda ^{2}}\int \mathrm{d}^{4}x%
\sqrt{g}\left( \frac{c_{1}}{\Lambda ^{4}}R_{\mu \nu }R^{\mu \nu }+\frac{c_{2}%
}{\Lambda ^{4}}R^{2}+\frac{c_{3}}{\Lambda ^{2}}R+c_{4}\right)
\end{eqnarray}%
The bare parameters $c_{i}$ are expressed in terms of the finite couplings $%
c_{i}^{r}\left( \mu \right) $ renormalized at the scale $\mu $ according to%
\footnote{%
Here we use the renormalization scheme suitable for power counting
non-renormalizable theories described in detail in \cite{Buchler:2003vw}} 
\begin{eqnarray}
c_{i} &=&c_{i}^{r}\left( \mu \right) +\frac{\beta _{i}}{\varepsilon }, 
\notag \\
\beta _{1} &=&-\frac{\Lambda ^{6}}{120\left( 4\pi \alpha \right) ^{2}}%
,~~\beta _{2}=-\frac{\Lambda ^{6}}{240\left( 4\pi \alpha \right) ^{2}}%
,~~\beta _{i>2}=0  \label{beta functions}
\end{eqnarray}%
and their renormalization scale dependence is determined in terms of the
coefficients $\beta _{i}$ as 
\begin{equation}
c_{i}^{r}\left( \mu ^{\prime }\right) =c_{i}^{r}\left( \mu \right) +2\beta
_{i}\ln \left( \frac{\mu ^{\prime }}{\mu }\right) .  \label{coupling running}
\end{equation}%
The natural values of $c_{i}^{r}\left( \mu \right) $ is of the order $%
O\left( 1\right) $. Therefore, in order to avoid incommensurable effects
stemming from the counterterms and from the loops, we expect $2\beta
_{j}=O\left( 1\right) $ and thus the natural value of the scale $\Lambda $ is%
\begin{equation}
\Lambda \simeq 2\left( 4\pi \alpha \right) ^{1/3}\simeq 4.6\times \alpha
^{1/3}.
\end{equation}%
Thus the consistency of the loop expansion requires that the scale $\alpha
^{1/3}$ which controls the strength of non-linearities in the basic
Lagrangian is roughly of the same order as the scale $\Lambda $ which
controls the systematic expansion of the quantum corrections.

\section{The scattering amplitudes at one loop and the soft theorem}

Using the results of the previous section we can write the renormalized
(i.e. finite) one-loop $S-$matrix in the form%
\begin{equation}
\mathcal{T}^{\mathrm{tree}}\left[ \phi \right] +\mathcal{T}^{\mathrm{1-loop}}%
\left[ \phi \right] =S_{0}\left[ \phi \right] +S_{\mathrm{eff}}^{\mathrm{%
1-loop}}\left[ \phi _{cl}\right] +\int \mathrm{d}^{D}x\phi _{cl}%
\overleftarrow{\square }\phi  \label{1-loop T matrix}
\end{equation}%
where we denoted as $S_{\mathrm{eff}}^{\mathrm{1-loop}}\left[ \phi _{cl}%
\right] $ the nonlocal one-loop effective action given by 
\begin{equation}
S_{\mathrm{eff}}^{\mathrm{1-loop}}\left[ \phi _{cl}\right] =S_{b}\left[ \phi
_{cl}\right] +S_{CT}^{\mathrm{1-loop}}\left[ \phi _{cl}\right] +\Gamma _{%
\mathrm{loop}}^{\mathrm{1-loop}}\left[ \phi _{cl}\right] ,
\label{effective action 1-loop}
\end{equation}%
with%
\begin{equation}
\Gamma _{\mathrm{loop}}^{\mathrm{1-loop}}\left[ \phi _{cl}\right] =\frac{%
\mathrm{i}}{2}\mathrm{Tr}\ln \frac{\delta ^{2}S_{b}\left[ \phi _{cl}\right] 
}{\delta \phi _{cl}\delta \phi _{c}}.  \label{Gamma 1-loop}
\end{equation}%
Note that $\phi _{cl}$ is still determined by (\ref{phi_cl definition}). Let
us now change for a moment the definition of $\phi _{cl}$ according to%
\begin{equation}
\phi _{cl}=\phi +\frac{1}{\square }\frac{\delta S_{\mathrm{eff}}^{\mathrm{%
1-loop,~int}}\left[ \phi _{cl}\right] }{\delta \phi _{cl}},
\label{redefinition of phi_cl}
\end{equation}%
where $S_{\mathrm{eff}}^{\mathrm{1-loop,~int}}$ is the interaction part of
the effective one-loop action $S_{\mathrm{eff}}^{\mathrm{1-loop}}$. Then we
can compare (\ref{1-loop T matrix}) and (\ref{redefinition of phi_cl}) with
the tree-level formulas (\ref{T^tree}) and (\ref{phi_cl definition}). Note
that the latter correspond to amplitudes given as a sum of the tree graphs
built from free propagators and vertices derived form the basic interaction
action $S_{b}^{\mathrm{int}}$. Therefore we can conclude, that the modified
prescription (\ref{1-loop T matrix}) and (\ref{redefinition of phi_cl})
corresponds to the scattering amplitudes constructed as the sum of the tree
graphs built from free propagators and (generally nonlocal) vertices derived
form the interaction part of the nonlocal effective action $S_{\mathrm{eff}%
}^{\mathrm{1-loop}}\left[ \phi _{cl}\right] $. \ Note, that these modified
one-loop amplitudes differ from the original prescription because the latter
allows at most one vertex from $S_{CT}^{\mathrm{1-loop}}\left[ \phi _{cl}%
\right] +\Gamma _{\mathrm{loop}}^{\mathrm{1-loop}}\left[ \phi _{cl}\right] $
in each graph, while the former has no such a constraint.

As we have proved in the previous section, the effective action $S_{\mathrm{%
eff}}^{\mathrm{1-loop}}\left[ \phi _{cl}\right] $ is invariant\footnote{%
Or at least can be made invariant provided we use invariant regularization
as DR and if we allow only invariant finite counterterms in $%
S_{CT,~~finite}^{1-loop}$, which we here tacitly assume.} with respect to
the hidden Special Galileon symmetry. Moreover, the effective action $S_{%
\mathrm{eff}}^{\mathrm{1-loop}}\left[ \phi _{cl}\right] $ contains only even
verteices, especially there is no cubic vertex present. Therefore, using the
general theorem connecting the symmetry of the action and the soft behavior
of the tree-level scattering amplitudes \cite{Cheung:2016drk}, we can
conclude that summing up all the tree graphs constructed with use of $S_{%
\mathrm{eff}}^{\mathrm{1-loop}}\left[ \phi _{cl}\right] $ we get amplitudes
with enhanced $O\left( p^{3}\right) $ soft limit, i.e. the enhanced soft
behavior is preserved also for the modified one-loop amplitudes.

Originally we were interested in the amplitudes given by graphs with index $%
i_{CT}^{\Gamma }\leq D+2$. These correspod to a subset of graphs
contributing to the modified one-loop scattering amplitudes, which can be
identified by counting the powers of the scale $\Lambda $ introduced in (\ref%
{CT infinite}) and (\ref{CT finite}), and keeping the contributions up to
and including the order $O\left( \Lambda ^{-D-2}\right) $. Here we have to
treat the vertices stemming form $\Gamma _{\mathrm{loop}}^{\mathrm{1-loop}}%
\left[ \phi _{cl}\right] $ as $O\left( \Lambda ^{-D+2}\right) $ (or with $%
i_{CT}^{\Gamma }=D+2$).

Because the contributions of the graphs with different $i_{CT}^{\Gamma }$
cannot cancell each other due to the different degree of homogeneity in
momenta, this above soft theorem remains true also when we restrict
ourselves to the contributions with $i_{CT}^{\Gamma }\leq D+2$ only. In this
sense the soft theorem for the Special Galileon is valid also at the
one-loop level.

\section{Special Galileon amplitudes at one loop - explicit example\label%
{section amplitudes}}

Let us illustrate the above considerations using explicit expressions for
the one-loop four point amplitude in $D=4$. For definiteness we will use the
\textquotedblleft plus\textquotedblright\ branch of the Special Galileon
with the basic Lagrangian (\ref{sGal1}). The results for the
\textquotedblleft minus\textquotedblright\ branch can be obtained by
appropriate analytic continuation in the parameter $\alpha $.

The contribution $i_{\Gamma }=0~$corresponds to the basic Lagrangian and
stemms from the four-point vertex\footnote{%
Note that we have to rescale the couplings (\ref{sGal1}) by a factor $-1/4!$
in order to canonically normalize the kinetic term.} 
\begin{equation}
\mathcal{L}_{b}^{\left( 4\right) }=\frac{1}{24\alpha ^{2}}\phi \varepsilon
^{\mu _{1}\mu _{2}\mu _{3}\mu }\varepsilon ^{\nu _{1}\nu _{2}\nu _{3}\nu
}\eta _{\mu \nu }\prod_{i=1}^{3}\partial _{\mu _{i}}\partial _{\nu _{i}}\phi
.
\end{equation}%
The corresponding off-shell vertex in the momentum representation is%
\begin{equation}
V^{\left( 4\right) }\left( p_{1},p_{2},p_{3},p_{4}\right) =-\frac{1}{\alpha
^{2}}G\left( p_{1},p_{2},p_{3}\right) ,  \label{4pt vertex}
\end{equation}%
where $G\left( p_{1},p_{2},p_{3}\right) $ is the Gramm determinant of the
momenta $p_{1}$, $p_{2}$ and $p_{3}$. The on-shell amplitude reads then%
\begin{equation}
A^{i_{\Gamma }=0}=-\frac{1}{4\alpha ^{2}}stu  \label{basic tree amplitude}
\end{equation}%
where $s$, $t$ and $u$ are the usual Mandelstamm variables%
\begin{eqnarray}
s &=&\left( p_{1}+p_{2}\right) ^{2}=\left( p_{3}+p_{4}\right) ^{2}  \notag \\
t &=&\left( p_{1}+p_{3}\right) ^{2}=\left( p_{2}+p_{4}\right) ^{2}  \notag \\
u &=&\left( p_{1}+p_{4}\right) ^{2}=\left( p_{2}+p_{3}\right) ^{2}
\label{Mandelstamm variables}
\end{eqnarray}%
and all the momenta are treated as outgoing. The $O\left( p_{i}^{3}\right) $
soft behavior of the amplitude $A^{i_{\Gamma }=0}$ is manifest. The pure
one-loop contributions corresponds to the bubble graphs with two vertices (%
\ref{4pt vertex}), explicitly%
\begin{equation}
A_{\mathrm{1-loop}}^{i_{\Gamma }=6}=\frac{1}{8}\sum_{\sigma \in S_{4}}A_{%
\mathrm{bubble}}\left( p_{\sigma \left( 1\right) },p_{\sigma \left( 2\right)
,}p_{\sigma \left( 3\right) ,}p_{\sigma \left( 4\right) }\right)
\label{one loop complet}
\end{equation}%
where%
\begin{eqnarray}
&&A_{\mathrm{bubble}}\left( p_{i},p_{j},p_{k},p_{l}\right) =\frac{1}{2}%
\left( \frac{1}{\alpha ^{2}}\right) ^{2}\left( p_{i}\cdot p_{j}\right)
\left( p_{k}\cdot p_{l}\right)  \notag  \label{one loop partial} \\
&&\times \int \frac{\mathrm{d}^{d}l}{\left( 2\pi \right) ^{d}}\frac{[2\left(
p_{i}\cdot q_{ij}\right) \left( p_{j}\cdot q_{ij}\right) -q_{ij}^{2}\left(
p_{i}\cdot p_{j}\right) ][2\left( p_{k}\cdot q_{kl}\right) \left( p_{l}\cdot
q_{kl}\right) -q_{kl}^{2}\left( p_{k}\cdot p_{l}\right) ]}{(q_{ij}^{2}+%
\mathrm{i}0)(q_{kl}^{2}+\mathrm{i}0)},  \notag \\
&&
\end{eqnarray}%
and where $q_{mn}=l+\frac{1}{2}\left( p_{m}+p_{n}\right) $. For further
convenience, let us also introduce the notation%
\begin{equation}
s_{ij}=s_{ji}=\left( p_{i}+p_{j}\right) ^{2}.
\end{equation}%
The result of the loop integration has the form%
\begin{equation}
A_{\mathrm{bubble}}\left( p_{i},p_{j},p_{k},p_{l}\right) =P\left(
s_{ij},s_{ik},s_{il};\varepsilon \right) B\left( s_{ij};\varepsilon \right)
\label{one loop partial result}
\end{equation}%
where $\varepsilon =2-d/2$ , the function $B\left( s_{ij}\right) $ is the
scalar two-point function given as 
\begin{eqnarray}
B\left( s_{ij}\right) &=&-\mathrm{i}\int \frac{\mathrm{d}^{d}l}{\left( 2\pi
\right) ^{d}}\frac{1}{[q_{ij}^{2}+\mathrm{i}0][q_{kl}^{2}+\mathrm{i}0]}=%
\frac{\mu ^{-2\varepsilon }}{\left( 4\pi \right) ^{2-\varepsilon }}\frac{%
\Gamma \left( \varepsilon \right) \Gamma \left( 1-\varepsilon \right) ^{2}}{%
\Gamma \left( 2-2\varepsilon \right) }\left( -\frac{s_{ij}}{\mu ^{2}}\right)
^{-\varepsilon }  \notag \\
&=&\frac{\mu ^{-2\varepsilon }}{\left( 4\pi \right) ^{2}}\left[ \frac{1}{%
\varepsilon }-\gamma +\ln 4\pi +2-\ln \left( -\frac{s_{ij}}{\mu ^{2}}\right)
+O\left( \varepsilon \right) \right] ,  \label{scalar bubble}
\end{eqnarray}%
and $P$ is the following polynomial%
\begin{equation}
P\left( s_{ij},s_{ik},s_{il};\varepsilon \right) =\frac{1}{512\alpha
^{4}\left( d^{2}-1\right) }s_{ij}^{4}\left[ s_{ij}^{2}d\left( d-2\right)
-8s_{ik}s_{il}\right] .
\end{equation}%
Finally we get for the bubble%
\begin{eqnarray}
A_{\mathrm{bubble}}\left( p_{i},p_{j},p_{k},p_{l}\right) &=&\frac{\mu
^{-2\varepsilon }}{\left( 4\pi \right) ^{2}}\frac{s_{ij}^{4}}{1920\alpha ^{4}%
}  \notag \\
&&\times \left\{ \left( s_{ij}^{2}+s_{ik}^{2}+s_{il}^{2}\right) \left[ \frac{%
1}{\varepsilon }+c_{\mathrm{bubble}}-\ln \left( -\frac{s_{ij}}{\mu ^{2}}%
\right) \right] -3s_{ij}^{2}\right\} .
\end{eqnarray}%
The actual value of the constant $c_{\mathrm{bubble}}$ depends on the
details of the dimensional regularization scheme. We use the t'Hooft-Veltman
scheme for the reduction of the tensor integrals, for which 
\begin{equation}
c_{\mathrm{bubble}}=-\gamma +\ln 4\pi +\frac{46}{15}.
\end{equation}%
The complete one-loop contribution to the amplitude is then 
\begin{eqnarray}
A_{\mathrm{1-loop}}^{i_{\Gamma }=6} &=&\frac{\mu ^{-2\varepsilon }}{\left(
4\pi \right) ^{2}}\frac{1}{1920\alpha ^{4}}\left\{ \left( \frac{1}{%
\varepsilon }+c_{\mathrm{bubble}}\right) \frac{1}{2}(s^{2}+t^{2}+u^{2})^{3}%
\right.  \notag \\
&&\left. -(s^{2}+t^{2}+u^{2})\left[ s^{4}\ln \left( -\frac{s}{\mu ^{2}}%
\right) +t^{4}\ln \left( -\frac{t}{\mu ^{2}}\right) +u^{4}\ln \left( -\frac{u%
}{\mu ^{2}}\right) \right] \right.  \notag \\
&&\left. \phantom{\frac{1}{2}}-3\left( s^{6}+t^{6}+u^{6}\right) \right\} .
\end{eqnarray}%
Let us now consider the contributions of the higher derivative counterterms.
At the level $i_{\Gamma }=2$ we have%
\begin{eqnarray}
c_{4}\frac{\alpha ^{2}}{\Lambda ^{2}}\sqrt{g} &=&c_{4}\frac{\alpha ^{2}}{%
\Lambda ^{2}}\left[ 1+\frac{1}{2\alpha ^{2}}\left\langle \partial \partial
\phi .\partial \partial \phi \right\rangle \right.  \notag \\
&&\left. -\frac{1}{4\alpha ^{4}}\left\langle \partial \partial \phi
.\partial \partial \phi .\partial \partial \phi .\partial \partial \phi
\right\rangle +\frac{1}{8\alpha ^{4}}\left\langle \partial \partial \phi
.\partial \partial \phi \right\rangle ^{2}+O\left( \phi ^{6}\right) \right] ,
\label{i=2 Lagrangian}
\end{eqnarray}%
where now the normal dot (and normal colon in what follows) means
contraction of adjacent indices with flat metric $\eta ^{\alpha \beta }$ and 
$\langle \cdot \rangle $ denotes a trace with respect to the same flat
metric. The effect of this Lagrangian is therefore twofold. The term
quadratic in the fields contributes to the off-shell two point function of
the field $\phi $; for this contribution we get explicitly%
\begin{equation}
\Sigma ^{i_{\Gamma }=2}\left( p^{2}\right) =-\frac{c_{4}}{\Lambda ^{2}}p^{4}.
\end{equation}%
Note however, that the derivative of $\Sigma ^{i_{\Gamma }=2}\left(
p^{2}\right) $ vanishes on shell and therefore there is no external leg
renormalization. The quartic terms in (\ref{i=2 Lagrangian}) are responsible
for the contact cotribution to the four point on-shell amplitude. We get%
\begin{equation}
A_{CT}^{i_{\Gamma }=2}=-\frac{c_{4}}{\Lambda ^{2}}\frac{1}{128\alpha ^{2}}%
\sum_{\sigma \in S_{4}}s_{\sigma \left( 1\right) \sigma \left( 2\right) }^{2}%
\left[ s_{\sigma \left( 1\right) \sigma \left( 4\right) }^{2}-\frac{1}{2}%
s_{\sigma \left( 1\right) \sigma \left( 2\right) }^{2}\right] =0
\end{equation}%
and thus there is no $i_{\Gamma }=2$ contribution to the four-point
amplitude. At the next level $i_{\Gamma }=4$ we have the following expansion
of the Lagrangian%
\begin{eqnarray}
c_{3}\frac{\alpha ^{2}}{\Lambda ^{4}}\sqrt{g}R &=&-\frac{c_{3}}{\Lambda ^{4}}%
\left[ \langle \partial \partial \partial \phi \colon \partial \partial
\partial \phi \rangle -\frac{3}{\alpha ^{2}}\langle \partial \partial
\partial \phi \colon \partial \partial \partial \phi .\partial \partial \phi
.\partial \partial \phi \rangle \right.  \notag \\
&&\left. +\frac{1}{2\alpha ^{2}}\langle \partial \partial \phi .\partial
\partial \phi \rangle \langle \partial \partial \partial \phi \colon
\partial \partial \partial \phi \rangle +O\left( \phi ^{6}\right) \right] .
\label{i=4 Lagrangian}
\end{eqnarray}%
Again, there is a contribution to the off-shell two point function, which,
however, has no effect to the on shell four point amplitude. The contact
contribution which stemms form the quartic term gives%
\begin{eqnarray}
A_{CT}^{i_{\Gamma }=4} &=&-\frac{c_{3}}{\Lambda ^{4}}\frac{3}{32\alpha ^{2}}%
\sum_{\sigma \in S_{4}}s_{\sigma \left( 1\right) \sigma \left( 2\right) }^{3}%
\left[ s_{\sigma \left( 1\right) \sigma \left( 4\right) }^{2}-\frac{1}{6}%
s_{\sigma \left( 1\right) \sigma \left( 2\right) }^{2}\right]  \notag \\
&=&\frac{c_{3}}{\Lambda ^{4}}\frac{1}{20\alpha ^{2}}\left(
s^{5}+t^{5}+u^{5}\right)
\end{eqnarray}%
Finally, for the $i_{\Gamma }=4$ counterterm contribution we have the
Lagrangian%
\begin{eqnarray}
\mu ^{-2\varepsilon }\frac{\alpha ^{2}}{\Lambda ^{6}}\sqrt{g}\left(
c_{1}R_{\mu \nu }R^{\mu \nu }+c_{2}R^{2}\right) &=&\frac{\mu ^{-2\varepsilon
}}{\Lambda ^{6}}\frac{1}{\alpha ^{2}}\left[ c_{1}\langle \partial \partial
\partial \phi \colon \partial \partial \partial \phi .\partial \partial
\partial \phi \colon \partial \partial \partial \phi \rangle \right.  \notag
\\
&&\left. +c_{2}\langle \partial \partial \partial \phi \colon \partial
\partial \partial \phi \rangle ^{2}+O\left( \phi ^{6}\right) \right] ,
\end{eqnarray}%
which yields the contact terms of the form%
\begin{eqnarray}
A_{CT}^{i_{\Gamma }=6} &=&\frac{\mu ^{-2\varepsilon }}{\Lambda ^{6}}\frac{1}{%
64\alpha ^{2}}\sum_{\sigma \in S_{4}}s_{\sigma \left( 1\right) \sigma \left(
2\right) }^{4}\left[ c_{1}s_{\sigma \left( 1\right) \sigma \left( 4\right)
}^{2}+c_{2}s_{\sigma \left( 1\right) \sigma \left( 2\right) }^{2}\right] 
\notag \\
&=&\frac{\mu ^{-2\varepsilon }}{\Lambda ^{6}}\frac{1}{8\alpha ^{2}}\left[ 
\frac{1}{4}c_{1}(s^{2}+t^{2}+u^{2})^{3}+\left( c_{2}-\frac{1}{2}c_{1}\right)
\left( s^{6}+t^{6}+u^{6}\right) \right]  \label{A_i=6}
\end{eqnarray}%
The infinite part of $A_{CT}^{i_{\Gamma }=6}$ then reads%
\begin{equation}
A_{CT,\infty }^{i_{\Gamma }=6}=-\frac{\mu ^{-2\varepsilon }}{\left( 4\pi
\right) ^{2}}\frac{1}{\varepsilon }\frac{1}{3840\alpha ^{4}}%
(s^{2}+t^{2}+u^{2})^{3},
\end{equation}%
and cancels the infinite part of the of the loop contribution $%
A_{1-loop}^{i_{\Gamma }=6}$. The finite part of $A_{CT}^{i_{\Gamma }=6}$can
be obtained from (\ref{A_i=6}) by the replacement $c_{i}\rightarrow
c_{i}^{r}\left( \mu \right) $. Finally we get\footnote{%
Here we used the identity 
\begin{equation*}
(s^{2}+t^{2}+u^{2})^{3}=-12s^{2}t^{2}u^{2}+4\left( s^{6}+t^{6}+u^{6}\right)
\end{equation*}%
in order to be able to compare the results of the section \ref{section
amplitudes}\ with existing literature.}%
\begin{eqnarray}
A^{i_{\Gamma }\leq 6} &=&-\frac{1}{4\alpha ^{2}}stu+\frac{1}{\Lambda ^{4}}%
\frac{c_{3}^{r}\left( \mu \right) }{20\alpha ^{2}}\left(
s^{5}+t^{5}+u^{5}\right) +\frac{1}{\Lambda ^{6}}\left( \frac{k_{1}^{r}\left(
\mu \right) }{8\alpha ^{2}}s^{2}t^{2}u^{2}+\frac{k_{2}^{r}\left( \mu \right) 
}{8\alpha ^{2}}\left( s^{6}+t^{6}+u^{6}\right) \right)  \notag \\
&&-\frac{1}{\left( 4\pi \right) ^{2}}\frac{1}{1920\alpha ^{4}}\left\{
(s^{2}+t^{2}+u^{2})\left[ s^{4}\ln \left( -\frac{s}{\mu ^{2}}\right)
+t^{4}\ln \left( -\frac{t}{\mu ^{2}}\right) +u^{4}\ln \left( -\frac{u}{\mu
^{2}}\right) \right] \right\} ,  \notag \\
&&
\end{eqnarray}%
where we abbreviated%
\begin{eqnarray}
k_{1}^{r}\left( \mu \right) &=&-3c_{1}^{r}\left( \mu \right) -\frac{\Lambda
^{6}}{\left( 4\pi \right) ^{2}}\frac{1}{40\alpha ^{2}}\left( -\gamma +\ln
4\pi +\frac{46}{15}\right)  \notag \\
k_{2}^{r}\left( \mu \right) &=&c_{2}^{r}\left( \mu \right) +\frac{1}{2}%
c_{1}^{r}\left( \mu \right) +\frac{\Lambda ^{6}}{\left( 4\pi \right) ^{2}}%
\frac{1}{120\alpha ^{2}}\left[ \left( -\gamma +\ln 4\pi +\frac{46}{15}%
\right) -\frac{3}{2}\right] .  \label{kr couplings}
\end{eqnarray}

Note, that the $O\left( p^{3}\right) $ soft behavior is manifest for all the
above components of the amplitude. In the case of the four-point amplitude
it is somewhat trivial statement due to the special four-particle kinematics
and due to the power counting of the individual contributions.

\section{Summary and discussion}

In this paper we have studied the issue of one-loop renormalization of the
Special Galileon $S-$matrix. First we calculated the UV divergent part $%
\Gamma _{\mathrm{loop}}^{\mathrm{1-loop}}\left[ \phi _{cl}\right] _{\infty }$
of the one-loop on-shell effective action and proved its invariance with
respect to the hidden Special Galileon symmetry for general space-time
dimension $D$. The key ingredient of the proof was the fact, that we were
able to express the action describing the quantum fluctuations in the
on-shell classical background in terms of the geometric building blocks
which were covariant with respect to the Special Galileon symmetry. We have
further found appropriate prescription for the transformation of the
fluctuating field with respect to the Special Galileon symmetry which
ensures the invariance of the fluctuation action and, as a consequence, also
the invariance of the complete one-loop on-shell effective action $\Gamma _{%
\mathrm{loop}}^{\mathrm{1-loop}}\left[ \phi _{cl}\right] $.

For $D=4$ case we constructed the complete set of independent higher
derivative counterterms relevant for the calculation of the scattering
amplitudes, up to and including graphs with index $i_{\Gamma }=D_{\Gamma
}-2E_{\Gamma }+2=6$. The resulting counterterm action can be expressed in a
manifestly invariant form with respect to the hidden Special Galileon
symmetry, namely%
\begin{equation}
S_{CT}^{\mathrm{1-loop}}\mathcal{=}\mu ^{-2\varepsilon }\int \mathrm{d}^{4}x%
\sqrt{g}\sum_{j=1}^{4}\frac{\alpha ^{2}}{\Lambda ^{i_{\mathcal{O}_{j}}}}c_{j}%
\mathcal{O}_{j}.
\end{equation}%
Here $\mathrm{d}^{4}x\sqrt{g}$ is invariant measure corresponding to the
effective metric%
\begin{equation}
g_{\mu \nu }=\eta _{\mu \nu }\pm \frac{1}{\alpha ^{2}}\partial _{\mu
}\partial _{\alpha }\phi \partial ^{\alpha }\partial _{\nu }\phi ,
\end{equation}%
and the invariant operators $\mathcal{O}_{j}$ are given in terms of the
extrinsic curvature tensor 
\begin{equation}
\mathcal{K}_{\alpha \mu \nu }=-\frac{1}{\alpha }\partial _{\alpha }\partial
_{\mu }\partial _{\nu }\phi
\end{equation}%
and in terms of the inverse effective metric $g^{\mu \nu }$as 
\begin{eqnarray}
\mathcal{O}_{1} &=&\langle \langle \mathcal{K}\boldsymbol{\colon }\mathcal{K}%
\boldsymbol{\cdot }\mathcal{K}\boldsymbol{\colon }\mathcal{K}\rangle \rangle
\notag \\
\mathcal{O}_{2} &=&\langle \langle \mathcal{K}\boldsymbol{\colon }\mathcal{K}%
\rangle \rangle ^{2}  \notag \\
\mathcal{O}_{3} &=&\langle \langle \mathcal{K}\boldsymbol{\colon }\mathcal{K}%
\rangle \rangle  \notag \\
\mathcal{O}_{4} &=&1.
\end{eqnarray}%
Here the bold dots and $\langle \langle \ldots \rangle \rangle $ mean
contractions and trace with respect to $g^{\mu \nu }$ respectively. These
operators form a complete basis of the on-shell counterterms relevant for
the renormalization of the $S-$matrix at one loop and $i_{\Gamma }\leq 6$.
We have identified the infinite parts of the bare coupling $c_{j}$ and
established the running of the corresponding renormalized couplings $%
c_{j}^{r}\left( \mu \right) $ with the renormalization scale $\mu $.

Note, that the action $S_{CT}^{\mathrm{1-loop}}$ generates only even
vertices (especially there is no cubic vertex present). As we have proved,
also the nonlocal part of the $i_{\Gamma }\leq 6$ on-shell effective action $%
S_{\mathrm{eff}}^{\mathrm{1-loop}}$ stemming form the loops is invariant
with respect to the hidden Special Galileon symmetry and contains only even
nonlocal vertices. Using the general theorem about the relation between
generalized polynomial shift symmetries and the soft behavior of the
scattering amplitudes we proved that the enhanced $O\left( p^{3}\right) $
soft limit of the amplitudes is preserved also at one-loop and $i_{\Gamma
}\leq 6$.

Let us stress, that the proof of the manifestly invariant form of $S_{CT}^{%
\mathrm{1-loop}}$ and of the nonlocal one-loop effective action $S_{\mathrm{%
eff}}^{\mathrm{1-loop}}$ heavily depends on the fact that only the on-shell
configurations satisfying the classical equation of motion derived form the
basic Lagrangians are relevant for the calculation of the on-shell
scattering amplitudes. Also, this classical equation of motion were used for
the elimination of the redundant operators and reduction of the basis of the
counterterms. Therefore, though the counterterm action $S_{CT}^{\mathrm{%
1-loop}}$ guaranties the finiteness of the on-shell scattering amplitudes,
it is not sufficient to make also the off-shell Green functions finite.

In order to renormalize also the one-loop off-shell Green functions, the
simple explicit calculation of the UV divergent part of the effective action
is not possible. In general we can expect that additional counterterms will
be needed, and that the simple structure of the higher order Lagrangians
will be lost. For instance, the most general counterterm action invariant
with respect to the hidden Special Galileon symmetry should be of the form (%
\ref{general counterterm action}), with additional invariant odd in the
field which vanish on-shell, namely\footnote{%
Here we present the formula for the \textquotedblleft
plus\textquotedblright\ branch of the transformation (\ref{sGal_transform}).
The \textquotedblleft minus\textquotedblright\ branch variant can be
obtained with the replacement $\alpha \rightarrow \mathrm{i}\alpha $.} 
\begin{equation}
\sigma =\frac{\alpha }{2\mathrm{i}}\ln \left[ \frac{\det \left( \eta +\frac{%
\mathrm{i}}{\alpha }\partial \partial \phi \right) }{\det \left( \eta -\frac{%
\mathrm{i}}{\alpha }\partial \partial \phi \right) }\right]
\end{equation}
Naively, at the level $i_{\Gamma }\leq 6$ we might expect the invariant
action of the form%
\begin{equation}
S_{CT}\mathcal{=}\mu ^{-2\varepsilon }\int \mathrm{d}^{4}x\sqrt{g}\sum_{j}%
\frac{\alpha ^{2}}{\Lambda ^{i_{\mathcal{O}_{j}}}}c_{j}\mathcal{O}_{j}
\label{naive action}
\end{equation}%
where the diffeomorphism invariant operators $\mathcal{O}_{j}$ are now
constructed as contractions (with respect to the effective metric $g^{\mu
\nu }$) of the extrinsic curvature tensor, its covariant derivatives and
covariant derivatives of the scalar $\sigma $. Schematically 
\begin{equation}
\mathcal{O}_{j}\sim \left( D^{n_{j}}\mathcal{K}^{m_{j}}\right) \left(
D^{r_{j}}\sigma \right) ^{k_{j}},
\end{equation}%
with $n_{j}+m_{j}+r_{j}k_{j}\leq 4$ (see e.g. the operators $\mathcal{O}%
_{5},\ldots ,\mathcal{O}_{8}$ discussed in section \ref{UV divergences at
one loop} and appendix \ref{counterterms}). To list a complete basis of such
off-shell independent operators is beyond the scope of this paper. However,
even if we were able to classify the operators $\mathcal{O}_{j}$ of the
above type, this is not the whole story. Note that the constants $c_{j}$ in (%
\ref{naive action}) can be freely replaced with arbitrary functions $%
c_{j}\left( \sigma \right) $ without changing the index of the corresponding
counterterms. Thus, the presence of the off-shell invariant $\sigma $
obscures the complete classification of the off-shell counterterm action $%
S_{CT}$. The situation is somewhat similar to the three-flavor chiral
perturbation theory with additional $U\left( 1\right) $ pseudoscalar
corresponding to $\eta ^{\prime }$ (cf. \cite{Gasser:1983yg, Gasser:1984gg}%
). In that case, the invariant Lagrangians is determined up to arbitrary
potentials which are functions of the $\eta ^{\prime }$ field. Note also,
that because the off-shell invariant $\sigma $ is odd in the field, the
off-shell basis should contain also the operators giving rise to the odd
off-shell vertices. However, the contributions of such vertices have to
cancel each other in the on-shell amplitudes.

Recently, a complementary classification of the higher order Special
Galileon Lagrangians in $D=4$ was developed using the coset construction
based on the Special Galileon symmetry algebra \cite%
{Garcia-Saenz:2019yok,Carrillo-Gonzalez:2019aao}. The basic building blocks
of this constructions are the invariant measure \textrm{d}$^{4}x\det \left(
E\right) $, the covariant derivative $\nabla _{a}\xi ^{a}$ of the Goldstone
field $\xi ^{a}$ corresponding to the linear part of the general Galileon
symmetry (\ref{general Galileon symmetry}) and the covariant derivative $%
\nabla _{c}\sigma _{ab}$ of the Goldstone field $\sigma _{ab}$ corresponding
to the hidden Special Galileon symmetry\footnote{%
The authors restrict themselves to the \textquotedblleft
plus\textquotedblright\ branch and their $\alpha $ parameter corresponds to
our $1/\alpha $.} (\ref{sGal_transform}). The Latin indices are then
contracted with the flat metric $\eta ^{ab}$ to build the invariants. Though
the direct comparison of these building blocks with our approach is
difficult, we have found the following correspondence with our geometrical
objects valid up to denoted higher orders in the field $\phi $%
\begin{eqnarray}
\det \left( E\right) &=&\sqrt{g}\left( 1-\frac{1}{8}\left( \frac{\sigma }{%
\alpha }\right) ^{2}+\frac{5}{768}\left( \frac{\sigma }{\alpha }\right) ^{4}-%
\frac{17}{92160}\left( \frac{\sigma }{\alpha }\right) ^{6}\right) +O\left(
\phi ^{8}\right)  \notag \\
\nabla _{a}\xi ^{a} &=&-\sigma \left( 1+\frac{1}{48}\left( \frac{\sigma }{%
\alpha }\right) ^{2}+\frac{1}{1920}\left( \frac{\sigma }{\alpha }\right)
^{4}\right) +O\left( \phi ^{7}\right)  \notag \\
\nabla _{c}\sigma _{ab} &=&\left( \mathcal{K}_{cab}-\frac{1}{4}\eta _{ab}%
\mathcal{K}_{c\mu \nu }g^{\mu \nu }\right) \left( 1+\frac{1}{32}\left( \frac{%
\sigma }{\alpha }\right) ^{2}\right) +O\left( \phi ^{5}\right) ,
\end{eqnarray}%
where%
\begin{equation*}
\mathcal{K}_{cab}=m_{a}^{\mu }m_{b}^{\nu }m_{c}^{\alpha }\mathcal{K}_{\alpha
\mu \nu },\,\,\,\mathcal{K}_{c\mu \nu }=m_{c}^{\alpha }\mathcal{K}_{\alpha
\mu \nu },
\end{equation*}%
and where 
\begin{equation}
m_{a}^{\mu }=\delta _{a}^{\mu }+\sum\limits_{n=1}^{\infty }\frac{\left(
1/2-n\right) _{n}}{n!}\frac{1}{\alpha ^{2n}}\left[ \left( \boldsymbol{%
\partial \partial \phi }\right) ^{2n}\right] _{a}^{\mu }
\end{equation}%
is the vielbein for the inverse effective metric $g^{\mu \nu }=\eta
^{ab}m_{a}^{\mu }m_{b}^{\nu }$ mentioned in the introduction (in a
particular gauge, see \cite{Novotny:2016jkh} for more details). We can thus
conclude, that on shell and up to the higher orders mentioned above%
\begin{equation}
\det \left( E\right) =\sqrt{g},~~\nabla _{a}\xi ^{a}=0\text{, ~}\nabla
_{c}\sigma _{ab}=\mathcal{K}_{cab}.
\end{equation}%
Therefore the building blocks for the on-shell higher derivative action are
the same as in our geometrical approach, at least for the vertices necessary
for the calculation of the 4pt, 5pt and 6pt scattering amplitudes at $%
i_{\Gamma }\leq 6$. This especially means, that there is no cubic and
quintic vertex relevant for the on-shell amplitudes. Though such vertices
might be present in the off-shell higher derivative action, e.g. in the
Lagrangian introduced in \cite{Carrillo-Gonzalez:2019aao} as 
\begin{equation}
\mathcal{L}_{\mathrm{odd}}^{i_{\Gamma }=2}\mathcal{=}\det \left( E\right)
\Delta \mathcal{L}^{\left( 0\right) }=\det \left( E\right) a_{1}\nabla
_{a}\xi ^{a}
\end{equation}%
their contribution have to vanish on shell. As a consequence, the leading
order $O(p^{10})$ 5pt amplitude proportional to $a_{1}$ as well as the $%
a_{1}^{2}$ contribution to the 4pt amplitude have to vanish on shell.

As an illustration of the one-loop renormalization of the Special Galileon
we have calculated the four-particle amplitude up to and including the
contributions with $i_{\Gamma }=6$. The result reads%
\begin{eqnarray}
A_{4}^{i_{\Gamma }\leq 6} &=&-\frac{1}{4\alpha ^{2}}stu+\frac{1}{\Lambda ^{4}%
}\frac{c_{3}^{r}\left( \mu \right) }{20\alpha ^{2}}\left(
s^{5}+t^{5}+u^{5}\right) +\frac{1}{\Lambda ^{6}}\left( \frac{k_{1}^{r}\left(
\mu \right) }{8\alpha ^{2}}s^{2}t^{2}u^{2}+\frac{k_{2}^{r}\left( \mu \right) 
}{8\alpha ^{2}}\left( s^{6}+t^{6}+u^{6}\right) \right)  \notag \\
&&-\frac{1}{\left( 4\pi \right) ^{2}}\frac{1}{1920\alpha ^{4}}\left\{
(s^{2}+t^{2}+u^{2})\left[ s^{4}\ln \left( -\frac{s}{\mu ^{2}}\right)
+t^{4}\ln \left( -\frac{t}{\mu ^{2}}\right) +u^{4}\ln \left( -\frac{u}{\mu
^{2}}\right) \right] \right\} ,  \notag \\
&&
\end{eqnarray}%
where $c_{j}^{r}\left( \mu \right) $ and $k_{j}^{r}\left( \mu \right) $ are
(linear combinations of) the couplings\footnote{%
See (\ref{kr couplings}) for explicit formulas.} renormalized at the scale $%
\mu $. The running of these couplings with $\mu $ given by (\ref{coupling
running}) and (\ref{beta functions}) ensures the manifest renormalization
scale independence of $A_{4}^{i_{\Gamma }\leq 6}$. This amplitude satisfies
also manifestly the soft theorem with $O\left( p^{3}\right) $ soft behavior.
Let us note that the polynomial part of $A_{4}^{i_{\Gamma }\leq 6}$ has been
determined independently in \cite{Carrillo-Gonzalez:2019aao} from the
Lagrangian constructed using the coset formalism with the same result up to
a redefinition of the couplings. Alternative determination of the polynomial
part was presented in \cite{Elvang:2018dco} using the soft bootstrap with
the conclusion, that consistency of the soft BCFW recursion for the
tree-level six-particle amplitude forces the constant $k_{2}^{r}\left( \mu
\right) $ to vanish. This conclusion was supported by the KLT double copy
construction of the amplitude. However, even if we set $k_{2}^{r}\left( \mu
\right) =0$ at some scale, due to the running of $k_{2}^{r}\left( \mu
\right) $ such a term is inevitably generated by the loop corrections at
other scale $\mu ^{\prime }$.

The next five-particle amplitude with $i_{\Gamma }\leq 6$ vanishes, since
there are no odd vertices in the one-loop on-shell effective action $S_{%
\mathrm{eff}}^{\mathrm{1-loop}}$%
\begin{equation}
A_{5}^{i_{\Gamma }\leq 6}=0
\end{equation}%
Note, that in \cite{Elvang:2018dco} it was constructed 5pt amplitude with $%
i_{\Gamma }=6$ consistent with the $O\left( p^{3}\right) $ soft behavior,
namely%
\begin{eqnarray}
A_{5}^{i_{\Gamma }=6} &=&\frac{c_{5}}{\Lambda ^{6}\alpha ^{3}}\varepsilon
_{\mu _{1}\mu _{2}\mu _{3}\mu _{4}}\sum_{\sigma \in S_{5}}p_{\sigma \left(
1\right) }^{\mu _{1}}p_{\sigma \left( 2\right) }^{\mu _{2}}p_{\sigma \left(
3\right) }^{\mu _{3}}p_{\sigma \left( 4\right) }^{\mu _{4}}  \notag \\
&&\times \left( p_{\sigma \left( 1\right) }\cdot p_{\sigma \left( 2\right)
}\right) \left( p_{\sigma \left( 2\right) }\cdot p_{\sigma \left( 3\right)
}\right) \left( p_{\sigma \left( 3\right) }\cdot p_{\sigma \left( 4\right)
}\right) \left( p_{\sigma \left( 4\right) }\cdot p_{\sigma \left( 5\right)
}\right) \left( p_{\sigma \left( 5\right) }\cdot p_{\sigma \left( 1\right)
}\right) .
\end{eqnarray}%
Though the soft bootstrap probe does not exclude such an amplitude, it was
established that it cannot be obtained by the KLT double copy construction.
In our approach, such an amplitude is excluded, since it cannot stem from
any operator invariant with respect to the Special Galileon symmetry.

Our results therefore support the conjecture, that symmetry based definition
of Special Galileon is in tension with the KLT double copy construction at
the higher orders as claimed in \cite{Carrillo-Gonzalez:2019aao}. However,
we did not confirm the violation of the interrelation between the Special
Galileon symmetry and the $O\left( p^{3}\right) $ soft behavior of the
amplitudes, since we proved that at least for $i_{\Gamma }\leq 6$ there does
not exist any cubic vertex relevant for the on-shell amplitudes.

\appendix

\section{Classification of the one loop on shell counterterms\label%
{counterterms}}

In this appendix we prove some statements concerning the elimination of
redundant on shell counterterms. Let us first remind the Codazzi equation
for the extrinsic curvature%
\begin{equation}
D_{\mu }K_{\nu \rho }^{a}-D_{\nu }K_{\mu \rho }^{a}=0
\end{equation}%
which is valid in our case due to the geometrical interpretation of the
Special Galileon field as an effective theory of the $D-$dimensional brane
in a flat $2D-$dimensional target space \cite{Novotny:2016jkh}. This can be
rewritten as a symmetry relation for the covariant derivative of the tensor $%
\mathcal{K}_{\alpha \mu \nu }=g_{\alpha \beta }m_{a}^{\beta }K_{\mu \nu
}^{a} $, where $m_{a}^{\beta }$ is the $D-$bein for the induced metric $%
g^{\mu \nu }$, as\footnote{%
Note that $g_{\alpha \beta }$ and $m_{a}^{\mu }$ are covariantly constant.}%
\begin{equation}
D_{\mu }\mathcal{K}_{\alpha \nu \rho }=D_{\nu }\mathcal{K}_{\alpha \mu \rho
}.  \label{Godazzi}
\end{equation}%
As a consequence,%
\begin{equation}
\left( D\boldsymbol{\cdot }\mathcal{K}\right) _{\nu \rho }=g^{\mu \alpha
}D_{\mu }\mathcal{K}_{\alpha \nu \rho }=g^{\mu \alpha }D_{\nu }\mathcal{K}%
_{\alpha \mu \rho }=D_{\nu }g^{\mu \alpha }\mathcal{K}_{\alpha \mu \rho }=0,
\end{equation}%
where the last equation holds on shell as a consequence of (\ref{EOM for K}%
). Therefore the operators 
\begin{eqnarray}
\mathcal{O}_{5} &=&\langle \langle \left( \mathcal{K}\boldsymbol{\colon }%
\mathcal{K}\right) \boldsymbol{\cdot }\left( D\boldsymbol{\cdot }\mathcal{K}%
\right) \mathcal{\rangle \rangle },  \notag \\
\mathcal{O}_{6} &=&\left( D\boldsymbol{\cdot }\mathcal{K}\right) \boldsymbol{%
\colon }\left( D\boldsymbol{\cdot }\mathcal{K}\right) ,
\end{eqnarray}%
vanish on shell. Similarly, with help of (\ref{Godazzi}) and (\ref{EOM for K}%
) we get on shell%
\begin{eqnarray}
\left( D\boldsymbol{\cdot }D\right) \mathcal{K}_{\alpha \beta \rho }
&=&g^{\mu \nu }D_{\mu }D_{\nu }\mathcal{K}_{\alpha \beta \rho }=g^{\mu \nu
}D_{\mu }D_{\beta }\mathcal{K}_{\alpha \nu \rho }=g^{\mu \nu }\left[ D_{\mu
},D_{\beta }\right] \mathcal{K}_{\alpha \nu \rho }+D_{\beta }g^{\mu \nu
}D_{\mu }\mathcal{K}_{\alpha \nu \rho }  \notag \\
&=&g^{\mu \nu }\left[ D_{\mu },D_{\beta }\right] \mathcal{K}_{\alpha \nu
\rho }=g^{\mu \nu }\left( R_{\alpha \beta \mu }^{\sigma }\mathcal{K}_{\sigma
\nu \rho }+R_{\nu \beta \mu }^{\sigma }\mathcal{K}_{\alpha \sigma \rho
}+R_{\rho \beta \mu }^{\sigma }\mathcal{K}_{\alpha \nu \sigma }\right) 
\notag \\
&=&g^{\mu \nu }g^{\sigma \kappa }\left( R_{\kappa \alpha \beta \mu }\mathcal{%
K}_{\sigma \nu \rho }+R_{\kappa \nu \beta \mu }\mathcal{K}_{\alpha \sigma
\rho }+R_{\kappa \rho \beta \mu }\mathcal{K}_{\alpha \nu \sigma }\right) .
\end{eqnarray}%
Now, using the Gauss equation for the Riemann tensor%
\begin{equation}
R_{\kappa \alpha \beta \mu }=\mathcal{K}_{\kappa \beta }\boldsymbol{\cdot }%
\mathcal{K}_{\alpha \mu }\mathcal{-K}_{\kappa \mu }\mathcal{\boldsymbol{%
\cdot }K}_{\alpha \beta },
\end{equation}%
and after some algebra, using the total symmetry of $\mathcal{K}_{\alpha \mu
\rho }$ and (\ref{EOM for K}) we have 
\begin{eqnarray}
\left( D\boldsymbol{\cdot }D\right) \mathcal{K}_{\alpha \beta \rho }
&=&2g^{\mu \nu }\mathcal{K}_{\nu \rho }\boldsymbol{\cdot }\mathcal{K}_{\beta
}\boldsymbol{\cdot }\mathcal{K}_{\alpha \mu }  \notag \\
&&\mathcal{-K}_{\rho }\boldsymbol{\colon }\mathcal{K\boldsymbol{\cdot }K}%
_{\alpha \beta }-\mathcal{K}_{\beta }\boldsymbol{\colon }\mathcal{K%
\boldsymbol{\cdot }K}_{\alpha \rho }-\mathcal{K}_{\alpha }\boldsymbol{\colon 
}\mathcal{K\boldsymbol{\cdot }K}_{\rho \beta }.
\end{eqnarray}%
Finally we get for the operator $\mathcal{O}_{7}$%
\begin{equation}
\mathcal{O}_{7}=\langle \langle \left( \mathcal{K}\boldsymbol{\colon }\left(
D\boldsymbol{\cdot }D\right) \mathcal{K}\right) \mathcal{\rangle \rangle }=2%
\mathcal{O}_{8}-3\mathcal{O}_{1}
\end{equation}%
where $\mathcal{O}_{1}$ and $\mathcal{O}_{8}$ are defined as%
\begin{eqnarray}
\mathcal{O}_{1} &=&\langle \langle \mathcal{K}\boldsymbol{\colon }\mathcal{K}%
\boldsymbol{\cdot }\mathcal{K}\boldsymbol{\colon }\mathcal{K}\rangle \rangle
=R_{\mu \nu }R^{\mu \nu }  \notag \\
\mathcal{O}_{8} &=&g^{\kappa \rho }g^{\gamma \beta }g^{\delta \alpha }g^{\mu
\nu }\mathcal{K}_{\kappa \gamma \delta }\mathcal{K}_{\nu \rho }\boldsymbol{%
\cdot }\mathcal{K}_{\beta }\boldsymbol{\cdot }\mathcal{K}_{\alpha \mu }.
\end{eqnarray}%
As far as the operator $\mathcal{O}_{8}$ term is concerned, let us remind,
that 
\begin{eqnarray}
R_{\kappa \alpha \beta \mu }R^{\kappa \alpha \beta \mu } &=&2\langle \langle 
\mathcal{K}\boldsymbol{\colon }\mathcal{K}\boldsymbol{\cdot }\mathcal{K}%
\boldsymbol{\colon }\mathcal{K}\rangle \rangle -2g^{\kappa \rho }g^{\gamma
\beta }g^{\delta \alpha }g^{\mu \nu }\mathcal{K}_{\kappa \gamma \delta }%
\mathcal{K}_{\nu \rho }\boldsymbol{\cdot }\mathcal{K}_{\beta }\boldsymbol{%
\cdot }\mathcal{K}_{\alpha \mu }  \notag \\
&=&2\mathcal{O}_{1}-2\mathcal{O}_{8},
\end{eqnarray}%
and thus the operators $\mathcal{O}_{7}$ and $\mathcal{O}_{8}$ can be
expressed in terms of the squares of the Riemann and Ricci tensor as%
\begin{eqnarray}
\mathcal{O}_{7} &=&-R_{\kappa \alpha \beta \mu }R^{\kappa \alpha \beta \mu
}-R_{\mu \nu }R^{\mu \nu }, \\
\mathcal{O}_{8} &=&R_{\mu \nu }R^{\mu \nu }-\frac{1}{2}R_{\kappa \alpha
\beta \mu }R^{\kappa \alpha \beta \mu }.
\end{eqnarray}%
Therefore up to a total derivative corresponding to the Gauss-Bonnet term $G$
(cf. (\ref{gauss bonnet})), these operators are linear combinations of $%
\mathcal{O}_{1}$ and $\mathcal{O}_{2}$ (see (\ref{O1}) and (\ref{O2})) 
\begin{eqnarray}
\mathcal{O}_{7} &=&-5R_{\mu \nu }R^{\mu \nu }+R^{2}-G=-5\mathcal{O}_{1}+%
\mathcal{O}_{2}-G, \\
\mathcal{O}_{8} &=&-R_{\mu \nu }R^{\mu \nu }+\frac{1}{2}R^{2}-\frac{1}{2}G=-%
\mathcal{O}_{1}+\frac{1}{2}\mathcal{O}_{2}-\frac{1}{2}G.
\end{eqnarray}

Let us now concentrate to the parity odd operators 
\begin{equation}
\mathcal{O}_{j}=\left( D^{n_{j}}\mathcal{K}^{m_{j}}\right) _{\mu _{1}\mu
_{2}\mu _{3}\mu _{4}}E^{\mu _{1}\mu _{2}\mu _{3}\mu _{4}},
\end{equation}%
where $n_{j}+m_{j}\leq 4$ and where%
\begin{equation}
E^{\mu _{1}\ldots \mu _{D}}=\frac{1}{\sqrt{g}}\varepsilon ^{\mu _{1}\ldots
\mu _{D}}.
\end{equation}
Note that, due to the symmetry of $\mathcal{K}_{\alpha \mu \nu }$, within
the tensor $\left( D^{n_{j}}\mathcal{K}^{m_{j}}\right) _{\mu _{1}\mu _{2}\mu
_{3}\mu _{4}}$each $\mathcal{K}$ can have at least one uncontracted index.
Also $g^{\mu \nu }\mathcal{K}_{\alpha \mu \nu }=0$ on shell, so that only
indices within different $\mathcal{K}$'s can be contracted.

There are no $i_{\Gamma }=2$ and $i_{\Gamma }=4$ such operators, since there
are not enough contractions to satisfy he above requirements. For instance
for $i_{\Gamma }=4$ the following possible operator with $n_{j}=0$, $m_{j}=2$
vanishes 
\begin{equation*}
\mathcal{O}_{9}^{i_{\Gamma }=4}=\left( \mathcal{K}_{\mu _{1}\mu _{2}}%
\boldsymbol{\cdot }\mathcal{K}_{\mu _{3}\mu _{4}}\right) E^{\mu _{1}\mu
_{2}\mu _{3}\mu _{4}}=0,
\end{equation*}%
since we have to contract the extrinsic curvature tensors $\mathcal{K}%
_{\alpha \mu \nu }$ once, however the result of this contraction is
symmetric in $\mu _{1}\mu _{2}$ $\ $and $\mu _{3}\mu _{4}$. Similarly,
because $D\boldsymbol{\cdot }\mathcal{K}_{\mu \nu }=0$ on shell, the
possible operator with $n_{j}=1$, $m_{j}=1$ 
\begin{equation*}
\mathcal{O}_{10}^{i_{\Gamma }=4}=D_{\mu _{1}}\mathcal{K}_{\mu _{2}\mu
_{3}\mu _{4}}E^{\mu _{1}\mu _{2}\mu _{3}\mu _{4}}=0.
\end{equation*}%
At the next level\footnote{%
Note that the operators with $n_{j}+3m_{j}$ odd cannot be constructed,
therefore there are no $i_{\Gamma }=5$ parity odd operators.} $i_{\Gamma }=6$
we have seemingly two operators, namely 
\begin{eqnarray*}
\mathcal{O}_{11}^{i_{\Gamma }=6} &=&g^{\alpha \beta }\mathcal{K}_{\alpha \mu
_{1}}\boldsymbol{\cdot }\mathcal{K}_{\mu _{2}}\boldsymbol{\cdot }\mathcal{K}%
_{\mu _{3}}\boldsymbol{\cdot }\mathcal{K}_{\mu _{4}\beta }E^{\mu _{1}\mu
_{2}\mu _{3}\mu _{4}} \\
\mathcal{O}_{12}^{i_{\Gamma }=6} &=&\left( \mathcal{K}_{\mu _{1}}\boldsymbol{%
\colon }\mathcal{K}_{\mu _{2}}\right) \left( \mathcal{K}_{\mu _{3}}%
\boldsymbol{\colon }\mathcal{K}_{\mu _{4}}\right) E^{\mu _{1}\mu _{2}\mu
_{3}\mu _{4}}
\end{eqnarray*}%
which however both vanish due to the symmetry (cyclic symmetry and $\mu
_{1}\mu _{2}$ $\ $and $\mu _{3}\mu _{4}$ symmetry respectively) of the four
times contracted extrinsic curvature tensors. In the case $n_{j}=1$, $%
m_{j}=3 $ we need three contractions within the $D$ $\mathcal{K}^{3}$
building block, so we have only the following possibility 
\begin{equation*}
\mathcal{O}_{13}^{i_{\Gamma }=6}=g^{\alpha \beta }\left( D_{\mu _{1}}%
\mathcal{K}_{\alpha \mu _{2}}\right) \boldsymbol{\cdot }\mathcal{K}_{\mu
_{3}}\boldsymbol{\cdot }\mathcal{K}_{\mu _{4}\beta }E^{\mu _{1}\mu _{2}\mu
_{3}\mu _{4}}
\end{equation*}%
which however vanish due to (\ref{Godazzi}). For $n_{j}=2$ and $m_{j}=2$ we
need two contractions within $D^{2}$ $\mathcal{K}^{2}$, the only possibility
with one uncontracted index per $\mathcal{K}$ is then (modulo integration by
parts) 
\begin{equation*}
\mathcal{O}_{14}^{i_{\Gamma }=6}=\left( D_{\mu _{1}}\mathcal{K}_{\mu
_{2}}\right) \boldsymbol{\colon }\left( D_{\mu _{3}}\mathcal{K}_{\mu
_{4}}\right) E^{\mu _{1}\mu _{2}\mu _{3}\mu _{4}}
\end{equation*}%
but this operator vanishes using (\ref{Godazzi}). Finally, in the case $%
n_{j}=3$ and $m_{j}=1$ only one contraction is needed within $D^{3}$ $%
\mathcal{K}$, however because $\mathcal{K}_{\alpha \mu \nu }$ is totally
symmetric and traceless, the subsequent contraction with $E^{\mu _{1}\mu
_{2}\mu _{3}\mu _{4}}$ vanishes. To summarize, there is no parity odd
on-shell operator up to and including $i_{\Gamma }=6$.

\vskip1cm

\textbf{Acknowledgment} The authors thank Mariana Carrillo Gonzalez,
Riccardo Penco and Mark Trodden for correspondence. This work is supported
in part by Czech Science Foundation (Project No. GA\v{C}R 18-17224S), and by
Ministry of Education, Youth and Sports of the Czech Republic (Project No.
LTAUSA17069).\bigskip

\bibliographystyle{utphys}
\bibliography{sGal_one_loop_biblio}
{}

\end{document}